\documentclass{aa}
\usepackage[varg]{txfonts}
\usepackage{natbib}
\usepackage{graphicx}
\usepackage{multirow}
\usepackage{xcolor}
\usepackage[norefs,nocites]{refcheck}
\usepackage{catchfilebetweentags}
\usepackage[group-separator={,}]{siunitx}
\newcommand{\no}{\ensuremath{\square}\hspace{0.24em}}
\newcommand{\yes}{\makebox[0pt][l]{$\square$}\raisebox{.15ex}{\hspace{0.12em}$\checkmark$}}

\newcommand{\radec}[2]{\multirow{#1}{*}{\ExecuteMetaData[tables/table_pos_ra.tex]{#2ra}} & \multirow{#1}{*}{\ExecuteMetaData[tables/table_pos_dec.tex]{#2dec}}}

\newcommand{\pms}[1]{\ExecuteMetaData[tables/table_pm_ra.tex]{#1ra} & \ExecuteMetaData[tables/table_pm_dec.tex]{#1dec}}
  
\newcommand{\mpy}[1]{\ExecuteMetaData[tables/table_mpy_ra.tex]{#1ra} & \ExecuteMetaData[tables/table_mpy_dec.tex]{#1dec}}
  
\newcommand{\chisqrthree}[1]{\ExecuteMetaData[tables/table_chisqr3.tex]{#1}}

\newcommand{\sigmo}[1]{\ExecuteMetaData[tables/table_significant_motion.tex]{#1}}
\newcommand{\um}{$\mu$m }

\definecolor{darkgreen}{HTML}{00AA00}
 
\bibpunct{(}{)}{;}{a}{}{,} 
\begin{document}
\title{Multi-epoch, high spatial resolution observations of multiple T Tauri systems}
\subtitle{}
\author{
Gergely~Csépány\inst{1,2}
\and 
Mario~van~den~Ancker\inst{1}
\and
Péter~Ábrahám\inst{2}
\and
Rainer~Köhler\inst{4,5}
\and
Wolfgang~Brandner\inst{3}
\and
Felix~Hormuth\inst{3}
\and
Hector~Hiss\inst{3}
} 
\institute{European Southern Observatory, Karl-Schwarzschild-Str. 2, 85748 Garching bei M\"unchen, Germany
\and 
Konkoly Observatory, Research Centre for Astronomy and Earth Sciences, Hungarian Academy of Sciences, H-1121 Budapest, Konkoly Thege Mikl\'os \'ut 15-17, Hungary
\and 
Max-Planck-Institut für Astronomie,  K\"onigstuhl 17,  69117 Heidelberg, Germany
\and
Institut f\"ur Astro- und Teilchenphysik, Universit\"at Innsbruck, Technikerstr. 25/8, 6020 Innsbruck, Austria
\and
University of Vienna, Department of Astrophysics, T\"urkenschanzstr. 17 (Sternwarte), 1180 Vienna, Austria
}
\date{Received / Accepted }
\abstract%
{
In multiple pre-main-sequence systems the lifetime of circumstellar disks appears to be shorter than around single stars, and the actual dissipation process may depend on the binary parameters of the systems. 
}
{
We report high spatial resolution observations of multiple T Tauri systems at optical and infrared wavelengths. We determine if the components are gravitationally bound and orbital motion is visible, derive orbital parameters and investigate possible correlations between the binary parameters and disk states.
}
{
We selected 18 T Tau multiple systems (16 binary and two triple systems, yielding $16+2\times2=20$ binary pairs) in the Taurus-Auriga star forming region from the survey by \cite{1993AnA...278..129L}, with spectral types from K1 to M5 and separations from $0.22\arcsec$ (31~AU) to $5.8\arcsec$ (814~AU). We analysed data acquired in 2006--07 at Calar Alto using the AstraLux lucky imaging system, along with data from SPHERE and NACO at the VLT, and from the literature.
}
{
We found ten pairs to orbit each other, five pairs that may show orbital motion and five likely common proper motion pairs. We found no obvious correlation between the stellar parameters and binary configuration. The 10~$\mu$m infra-red excess varies between 0.1 and 7.2 magnitudes (similar to the distribution in single stars, where it is between 1.7 and 9.1), implying that the presence of the binary star does not greatly influence the emission from the inner disk. 
}
{
We have detected orbital motion in young T Tauri systems over a timescale of $\approx20$ years. Further observations with even longer temporal baseline will provide crucial information on the dynamics of these young stellar systems.
}
\maketitle

\section{Introduction}

The influence of the multiplicity on the evolution of the protoplanetary disk is an open question of astronomy \citep[see e.g.][]{2006ApJ...653L..57B,2012ApJ...745...19K}.  
Multiple systems are present abundantly among young stars, as a large fraction of stars form in binary or multiple systems, and, for example,  \citet{2006ApJ...640L..63L} has found that the single star fraction is only $\approx$40\% for G-type stars, rising to 70\% for late M-type stars. This shows that multiplicity might be a significant factor in the stellar evolution.
The young multiple systems are of special interest, because the low-mass evolutionary models at the early phases of stellar evolution are currently only poorly constrained by observations, and observing such systems can help to refine and calibrate those models \citep{2009IAUS..258..161S}. 

There is evidence that the age at which the star still has a disk, is shorter in multiple systems than around single stars \citep[see e.g.][]{2007ApJ...670.1337D,2006ApJ...653L..57B}. In addition, \citet{1995ApJ...439..288O} claimed that companions closer than 100 AU inhibit disk formation, based on a 1.3~mm continuum survey of 121 young stars. \citet{2005ApJ...631.1134A} found that among the 150 young stars in Taurus (including 62 multiple systems) the sub-millimeter flux densities (and thus the disk masses) are lower for binaries with a projected semi-major axes < 100 AU. 
\citet{2009ApJ...696L..84C} also found that circumstellar disk lifetimes are reduced in binaries with separation less than 10--100 AU. Although the disks of the components are usually assumed to be coeval, they may evolve with different pace. 
This differential disk dispersion is in the focus of recent studies \citep{2008ApJ...677..616P,2009ApJ...703.1511K,2009ApJ...696L..84C,2012ApJ...745...19K,2012A&A...540A..46D,2012IAUS..282..452D,2013A&A...554A..43D}. However, although these studies derived some correlations between the presence of a disk and binarity, they did not perform any detailed analysis of disk content and structure. Simulations combining viscous draining and X-ray photoevaporation suggest that in systems with separation below 100 AU it is the secondary disk which disappears first, while for wider systems this trend cannot be seen \citep[][and references therein]{2014prpl.conf..267R}. 
The actual disk removal process that causes the relatively short disk lifespan and the differential disk dispersion is unclear. The disk removal processes may consist of tidal effects in a binary or multiple systems which destabilizes the disk's structure or the disk may deplete over time by accretion onto the star. It is also possible that planet formation is responsible for clearing the disk, or the disk may suffer photoevaporation from the star. 

Which of these effects is the main driver for disk dissipation in multiple systems is still unclear. However, it is likely that the actual disk dissipation process depends on the binary parameters of the systems, such as the age, separation, mass ratios and multiplicity of the systems. Therefore, to determine which process plays a dominant role of the dissipation, we need to witness the differences in disk clearing and determine binary parameters in a sample large enough to make statistical conclusions. 

Obtaining orbital parameters for multiple T Tauri systems has been the focus of recent works \citep{2006AJ....132.2618S,2014AJ....147..157S,2015csss...18..237K} and although some closer binaries have well-constrained orbital solutions, many wider systems still lack proper time coverage of their orbits \citep[see e.g.][for the case of the T Tauri system]{2015A&A...578L...9C}. This can be alleviated by observing the same systems over a long period of time to obtain data that cover a significant fraction of the orbit. 

In this paper we report high spatial resolution observations in the optical and infra-red of a sample of 18 multiple T Tauri systems. Most of the systems have been first resolved about 20 years ago, therefore having a timeline on which we may be able to see orbital motion. One of our aims was to determine if the systems are gravitationally bound, and if they are, then derive the orbital parameters and see if we find any correlation between their binary configuration and disk states.

\section{Sample}

We study a sample consisting of 18 T Tauri multiple systems (comprising of 16 binary and two triple systems, which are treated as two binary pairs for each triplet, where the components B and C are both measured relative to A, the brightest star in the V or R bands, depending on available measurements; altogether $16+2\times2=20$ binary pairs) in the Taurus-Auriga star forming region. Our sample is based on 
\cite{1993AnA...278..129L}, in which they conducted a survey from September 1991 to October 1992 using speckle imaging at the 3.5m telescope at Calar Alto. They found 44 multiple systems out of the 104 observed young low-mass stellar systems, and their measurements serve as an astrometric and photometric epoch for our targets. 
Our selection criteria considered observability using the employed telescope and instrumentation, together with the available observation time to ensure that we only include stars for which the separation and relative brightness of the components allow a reliable detection. The selected sample covers spectral types from K1 to M5, with separations from $0.22''$ to $5.8''$. Most of the systems are well-studied stars, with multiple epochs available, but we also included a few systems which have not been extensively observed. Therefore our new observations present the first time in which these system have been spatially resolved since the pioneering work of \citet{1993AnA...278..129L}.
The detailed description of the systems is found in Appendix~\ref{sec:sample}.

The proper motions of the main stars that we used for the analysis in Section~\ref{sec:results} are obtained from the publicly available UCAC4 all-sky star catalogue \citep{2012yCat.1322....0Z} and for one star (FV Tau/c) from the USNO-B catalogue \citep{2003AJ....125..984M}.

\subsection{Naming scheme}

The naming of the companions is not evident when we compile a database from observations spanning several decades. Some companions may have been only resolved recently, while others were simply renamed. The rename usually was minor, such as 'HK Tau/c' to 'HK Tau B'. However, sometimes we have to be very careful, as in the case of 'FV Tau', where 'FV Tau' and 'FV Tau/c' denotes two binary systems with a distance of $\approx 12\arcsec$ from each other. We tried our best to match the different names with each other, and to make sure that one name indeed refers to the same physical object. 
The naming scheme we employed is based on that the optically brightest is A, the second brightest is B and so on, with historical considerations to stay compatible with the most popular nomenclature in the literature. The final naming scheme is listed in Table~\ref{table:spectraltypes}.

\subsection{Distances}
\label{sec:distances}
\citet{2005ApJ...619L.179L,2007ApJ...671..546L,2009ApJ...698..242T,2007ApJ...671.1813T,2012ApJ...747...18T} carried out a long VLBI campaign to measure the distance of the nearby star forming regions, in which they measured three sections of the Taurus region. Taking into account all distances measured in this region and their spread (which probably reflects the three-dimensional extent of the Taurus cloud), we adopted $140 \pm 21$~pc as the distance of the stars in our sample.

\begin{table*}
\setlength{\tabcolsep}{3pt}
\caption{Spectral types, extinction magnitudes, classification and naming scheme notes of the stars in our sample}
\label{table:spectraltypes}
\centering
\begin{tabular}{llllllllll}
\hline\hline
\multicolumn{2}{l}{System \& Comp.} & RA$^1$ & DEC$^1$ & SpT & $A_V$ (mag)$^2$ & T Tau & Disk & Ref.$^5$ & Notes \\
&&&&& & class$^3$ & detected$^4$ & \\
\hline
       LkCa 3 & Aa & \radec{4}{lkca3} & M2   & 0.4 (nr) & \multirow{4}{*}{III} & NA & 1 & \multirow{4}{5.2cm}{The Aa-Ab and Ba-Bb pairs are spatially unresolved, spectroscopic binaries.}\\
       LkCa 3 & Ab & & & M4   & 0.4 (nr) & & NA & 1 & \\
       LkCa 3 & Ba & & & K7   & 0.4 (nr) & & NA & 1 & \\
       LkCa 3 & Bb & & & M2.5 & 0.4 (nr) & & NA & 1 & \\[0.4em]
       DD Tau & A & \radec{2}{ddtau} & M3    & 2.10 & \multirow{2}{*}{II} & \yes & 2,11 & \\
       DD Tau & B & & & M3    & 2.90 & & \no & 2,11 & \\[0.4em]
       LkCa 7 & A & \radec{2}{lkca7} & M0   & 0.6 (nr) & \multirow{2}{*}{III} &  NA &3 & \\
       LkCa 7 & B & & & M3.5  & 0.6 (nr) & &  NA & 3 & \\[0.4em]
       FV Tau & A & \radec{2}{fvtau} & K5  & 5.3 & \multirow{2}{*}{II} &  \yes & 4,11 & \multirow{4}{5.2cm}{FV Tau and FV Tau/c are two binary systems at $\approx 12\arcsec$ distance. \\ FV Tau/c is also known as HBC 387. } \\
       FV Tau & B & & & K6   & 5.3 & &  \yes & 4,11 & \\[0.4em]
     FV Tau/c & A & \radec{2}{fvtauc} & M2.5  & 3.25 & \multirow{2}{*}{II} &  \no & 2,11 & \\
     FV Tau/c & B & & & M3.5  & 7.00 & &  \no & 2,11 & \\[0.4em]
       UX Tau & A & \radec{4}{uxtau} & K5    & 1.8 & II &  \yes & 5,11 & \multirow{4}{5.2cm}{The three most luminous components are usually called A, B and C. However, UX Tau B is a binary, and when it is resolved, the pair is called B-C and C is renamed to D \citep{2006AnA...459..909C}.} \\
       UX Tau & B & & & M2    & 0.26 & III &  \no & 6,11 & \\
       UX Tau & C & & & M5    & 0.57 & II &  \no & 6,11 & \\
       UX Tau & D & & & --    & -- &  &  NA &  & \\
               & & & & & & & \\[0.4em]
       FX Tau & A & \radec{2}{fxtau} & M1    & 2.0 & \multirow{2}{*}{II} &  \yes & 7,11 & \\
       FX Tau & B & & & M4    & 2.0 & &  \no & 7,11 & \\[0.4em]
       DK Tau & A & \radec{2}{dktau} & K9    & 1.3 & II &  \yes & 7,11 & \\
       DK Tau & B & & & M1    & 1.3 & II &  \no & 7,11 & \\[0.4em]
       XZ Tau & A & \radec{2}{xztau} & M2    & 1.40 & \multirow{2}{*}{II} & \multirow{2}{*}{\no} & 2,11 & \\
       XZ Tau & B & & & M3.5  & 1.35 & &  & 2,11 & \\[0.4em]
       HK Tau & A & \radec{2}{hktau} & M1    & 3.0 & \multirow{2}{*}{II} &  \yes & 7, 8,11  & \multirow{2}{5.2cm}{B is denoted as /c in    \citet{1995ApJS..101..117K}} \\
       HK Tau & B & & & M2    & 2.5 & &  \yes & 7, 8,11  & \\[0.4em]
     V710 Tau & A & \radec{2}{v710tau} & M0.5  & 1.80 & II &  \yes & 6,11 & Also known as LkH$\alpha$ 266  \\
     V710 Tau & B & & & M2.5  & 1.82 & II &  \no & 6,11 & or HBC 51 or HBC 395.\\[0.4em]
       UZ Tau & A & \radec{3}{uztau} & M2    & 0.55 & II &  \yes & 2,11 & \\
       UZ Tau & B & & & M3    & 1.75 & \multirow{2}{*}{II} &  \yes & 2,11 & \\
       UZ Tau & C & & & M1    & 1.00 & &  \yes & 6,11 & \\[0.4em]
       GH Tau & A & \radec{2}{ghtau} & M2    & 0.00 & \multirow{2}{*}{II} & \multirow{2}{*}{\no} & 2,11 & \\
       GH Tau & B & & & M2    & 0.50 & &  & 2,11 & \\[0.4em]
       HN Tau & A & \radec{2}{hntau} & K5    & 0.6 & \multirow{2}{*}{II} &  \yes & 3,11 & \\
       HN Tau & B & & & M4    & 0.9 & &  \no & 3,11 & \\[0.4em]
       HV Tau & A & \radec{2}{hvtau} & M2    & 1.95 (nr) & III &  \no & 6,11 & \multirow{2}{5.2cm}{AB is a close binary, (74mas, \citet{1996ApJ...469..890S}), unresolved in our observations. In the WISE database, the object J043835.47+261041.8 is labelled as HV Tau A, but photometry suggests that it is actually HV Tau C.} \\
       HV Tau & C & & & M0    & 1.95 (nr) & I? &  \yes & 10,11 & \\
        & & & &  & & & \\
        & & & & & & & \\
        & & & & & & & \\
        & & & & & & & \\[1em]
  V999 Tau & A & \radec{2}{v999tau} & M0.5  & 2.00 & \multirow{2}{*}{III} &  NA & 2 & Also known as LkH$\alpha$ 332 G2. \\
  V999 Tau & B & & & M2.5  & 3.30 & &  NA & 2 &  \\[0.4em]
  V1000 Tau & A & \radec{2}{v1000tau} & M1    & 6.1 (nr) & \multirow{2}{*}{III} &  NA & 13  & Also known as LkH$\alpha$ 332 G1.\\
  V1000 Tau & B & & & K7    & 6.1 (nr) & &  NA & 13 & \\[0.4em]
       RW Aur & A & \radec{2}{rwaur} & K1    & 0.39 & \multirow{2}{*}{II} &  \yes & 6,12 & \\
       RW Aur & B & & & K5    & 1.56 & &  \no & 6,12 & \\
\hline
\multicolumn{10}{l}{$^1$ RA, DEC: coordinate of the center of light, $^2$ nr: non-resolved observation}\\
\multicolumn{10}{l}{$^3$ Based on \citet{2010ApJS..186..111L}, who classified the stars using Spitzer data.}\\
\multicolumn{10}{l}{$^4$ Whether the presence of a disk is detected in radio observations, based on Ref.~11 and 12.}\\
\multicolumn{10}{l}{$^5$ References: 
(1) \citet{2013ApJ...773...40T}, 
(2) \citet{2003ApJ...583..334H}, 
(3) \citet{2012ApJ...744..121Y},
(4) \citet{2011ApJ...740...43S},}\\
\multicolumn{10}{l}{
(5) \citet{2010ApJ...717..441E},
(6) \citet{2001ApJ...556..265W},
(7) \citet{1999AA...351..954D},
(8) \citet{1998ApJ...502L..65S},
}\\ \multicolumn{10}{l}{
(9) \citet{2009ApJ...703.1964F},
(10) \citet{2010ApJ...712..112D},
(11) \citet{2012ApJ...751..115H},
(12) \citet{2006AnA...452..897C},
(13) \citet{2010ApJ...724..835W}
}\\
\hline
\end{tabular}
\end{table*}

\section{Observations}

\subsection{AstraLux Norte}

We observed the selected 18 multiple systems between November 2006 and November 2007, at Calar Alto using the 2.2m telescope and the Astralux Norte lucky imaging system \citep[for a detailed description of Astralux Norte and the reduction pipeline see][]{2008SPIE.7014E.138H}.
In the lucky imaging process we took 10,000 images per object and exposed for 30--50~ms for each image. Johnson I and SDSS $i'$ and $z'$ filters were used, since at that exposure time and telescope size, those are the optimal wavelengths (800--1000~nm) to minimize the effect of the atmospheric turbulence \citep{1965JOSA...55.1427F}. Then we measured the Strehl-ratios of the reference star in the images and selected the best 1--5\% of the frames, which were composed into a final image using shift-and-add technique with the Drizzle algorithm \citep{2002PASP..114..144F}. 
This method selects the images in which the effect of the atmospheric turbulence was the lowest (as the turbulences in the air cells at the given wavelength operate at a larger timespan than 30--50~ms), allowing us to reach the diffraction limit of the telescope. The typical FWHM of the primary stars in the AstraLux images range between 80 and 92~mas, depending on the filter.

\begin{figure*}
\centering
\includegraphics[width=0.32\textwidth]{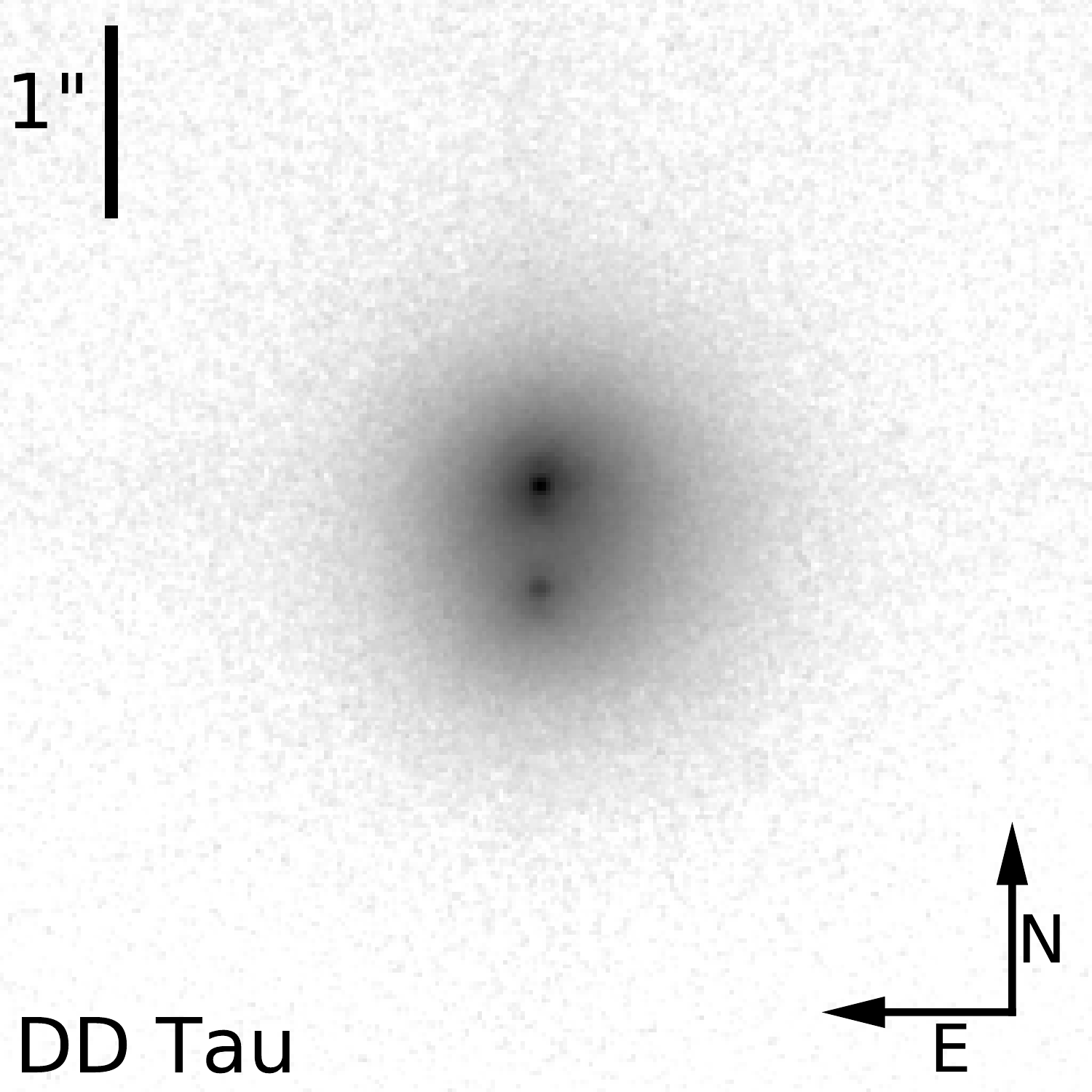}
\includegraphics[width=0.32\textwidth]{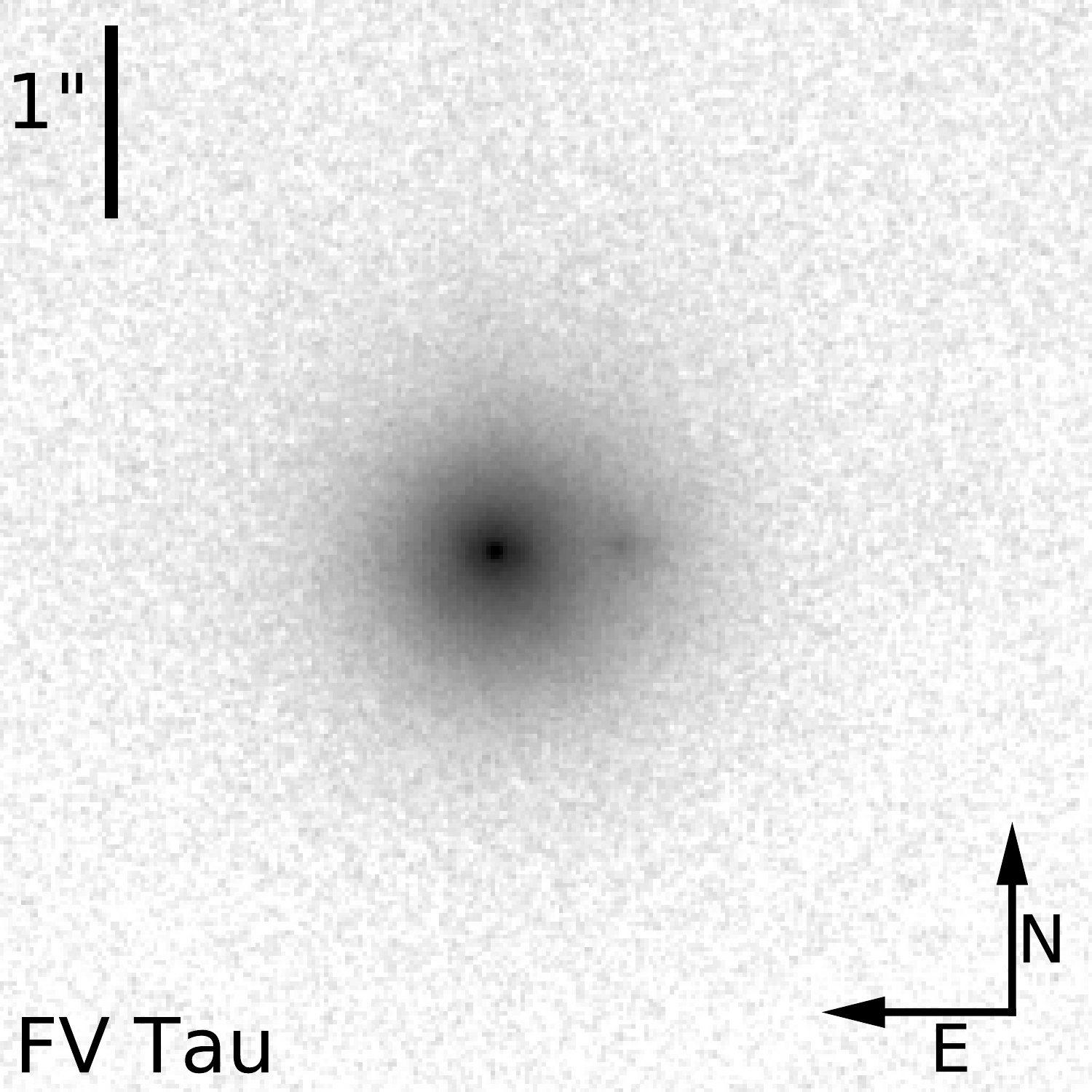}
\includegraphics[width=0.32\textwidth]{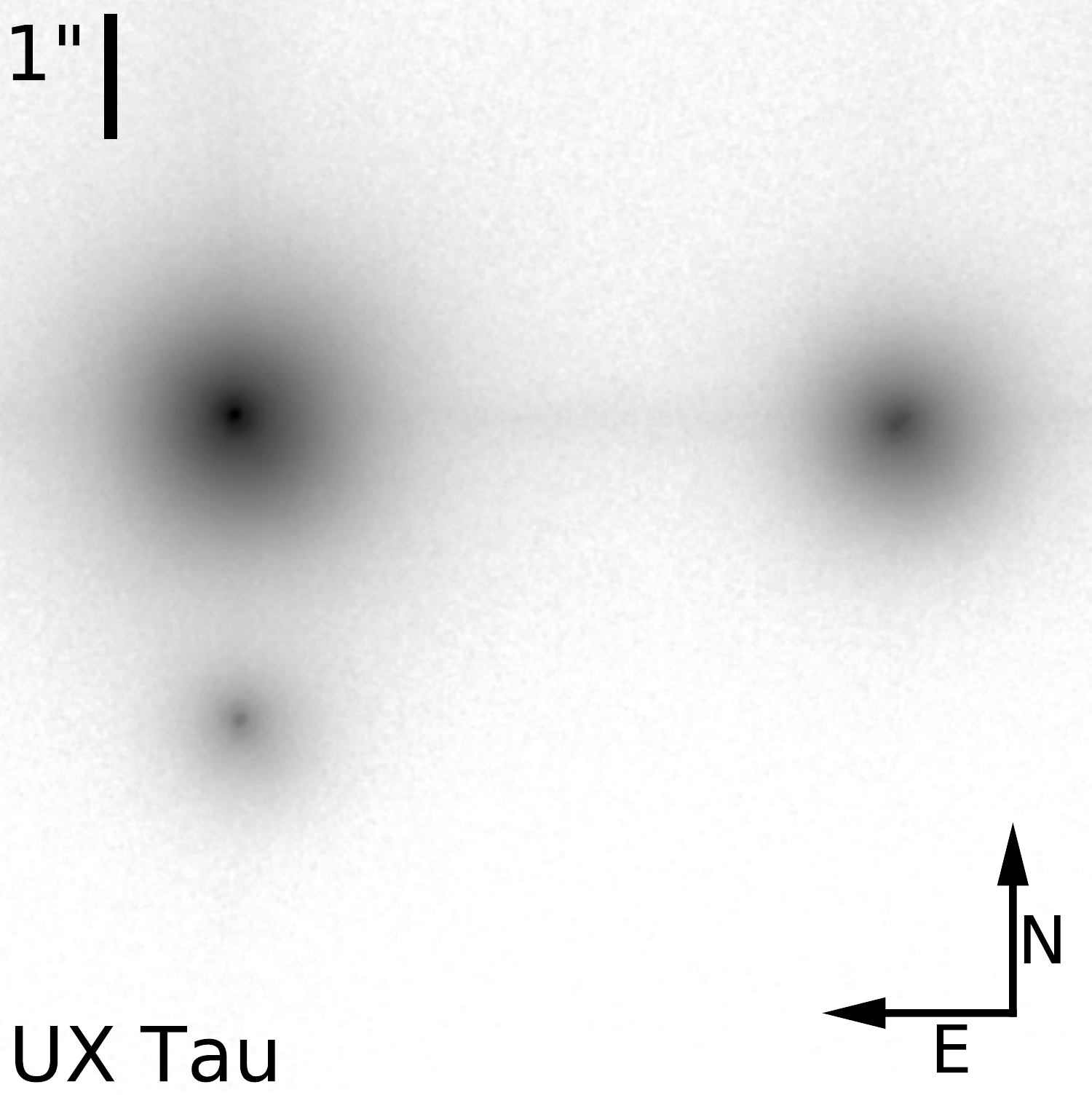}
\includegraphics[width=0.32\textwidth]{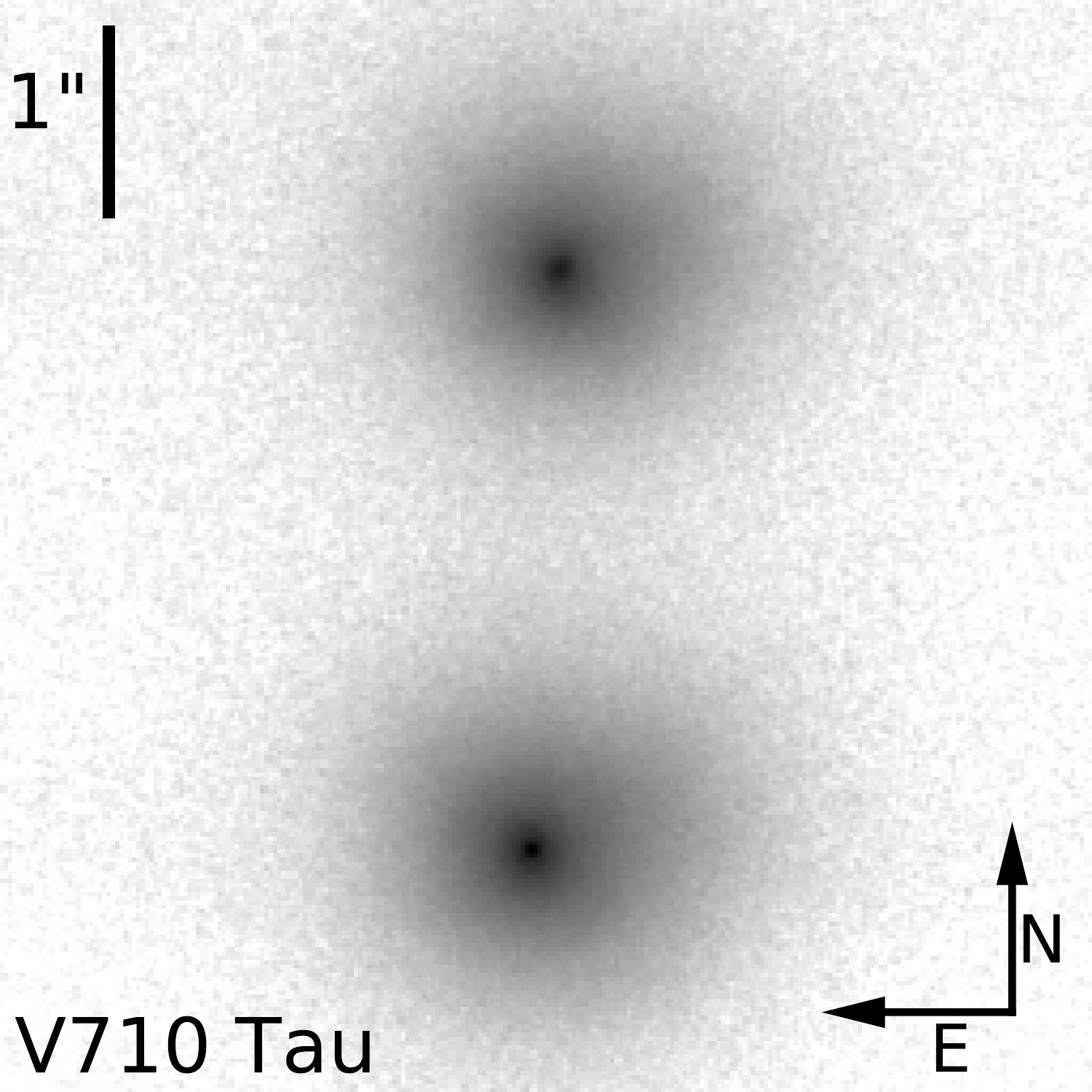}
\includegraphics[width=0.32\textwidth]{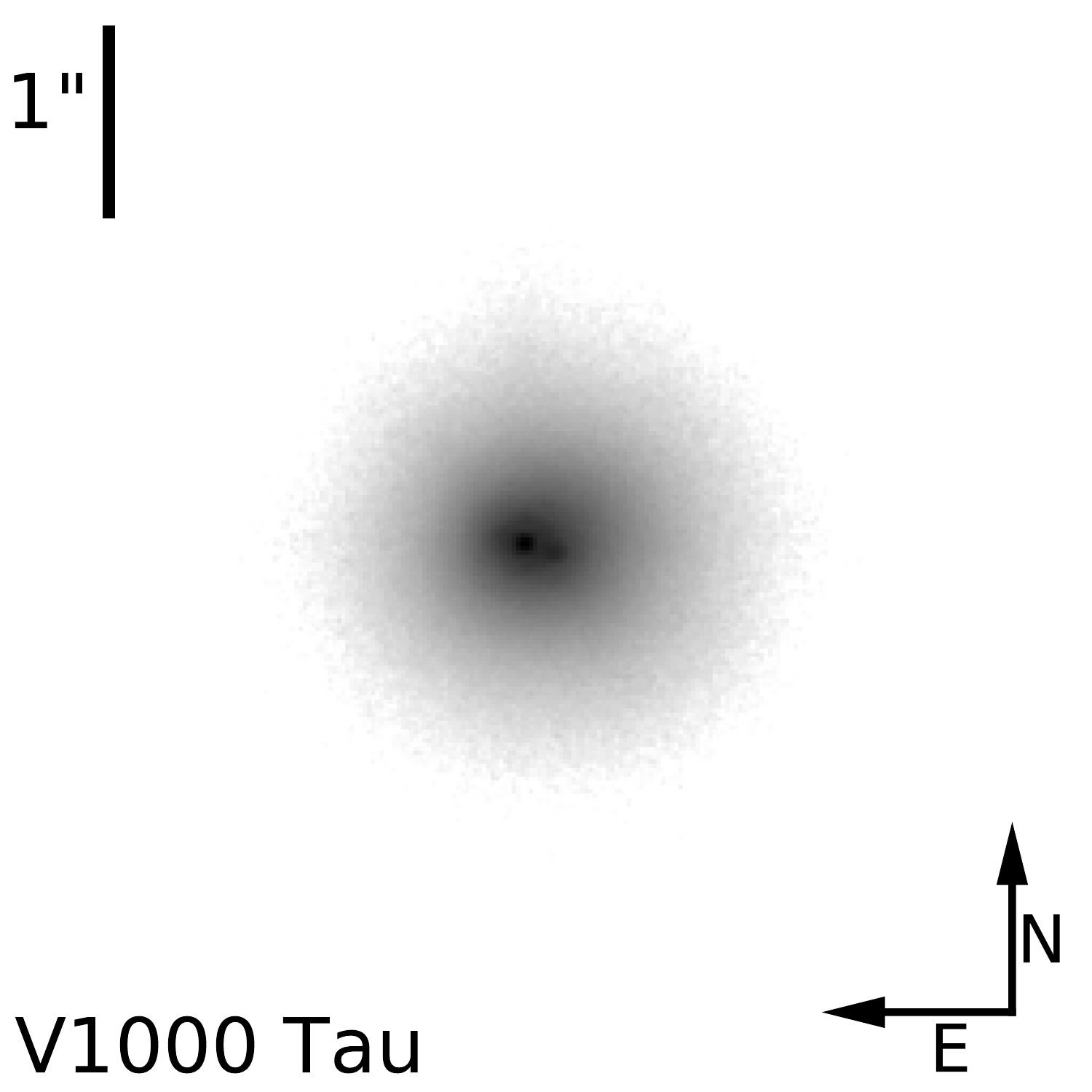}
\includegraphics[width=0.32\textwidth]{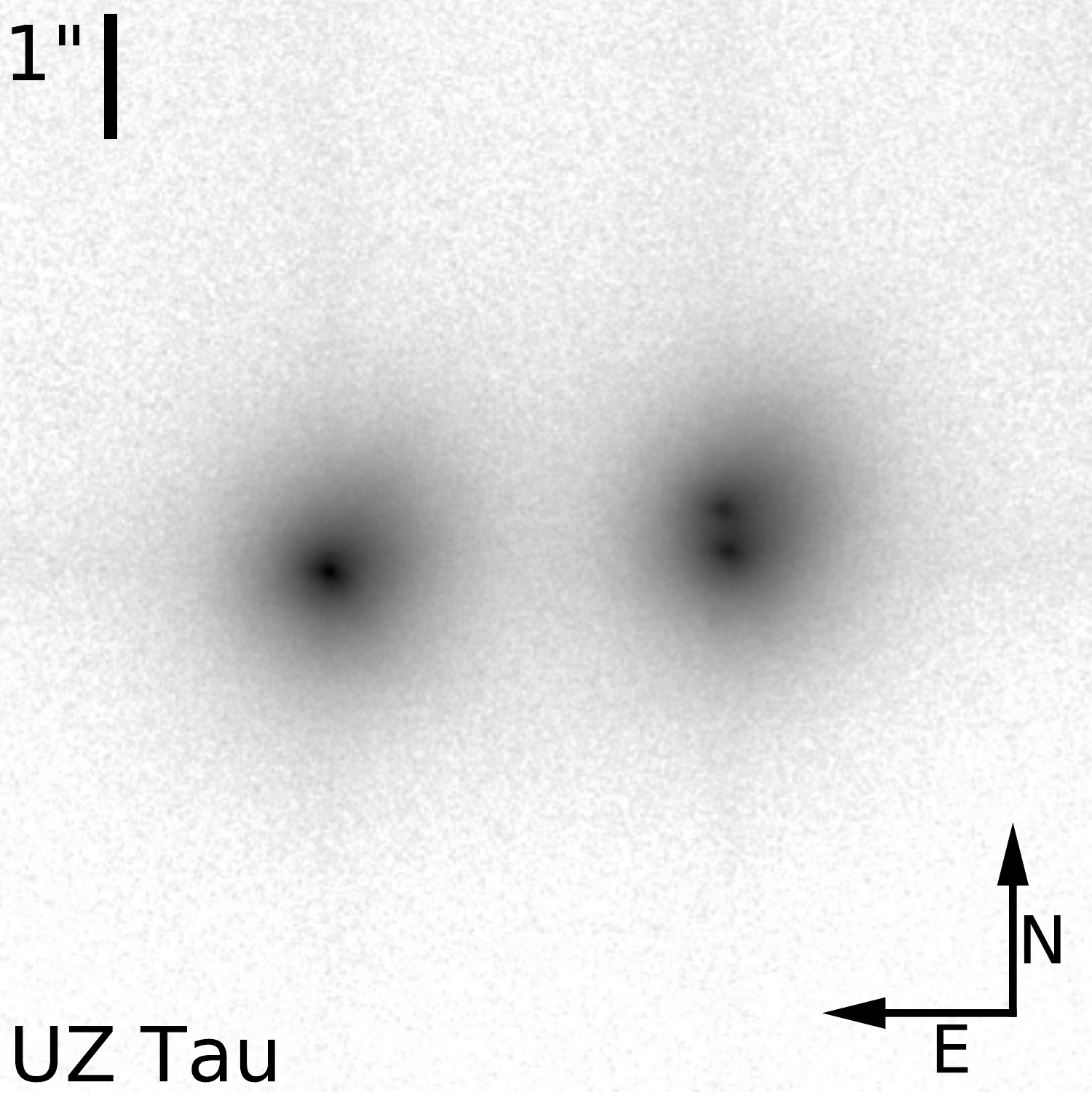}
\caption{AstraLux images from the sample showing a pair of similar brightness, a pair of a bright and a faint component, a wide, a tight pair and the two triple systems. The halos due to the lucky imaging addition are visible around the clearly resolved central cores of the stars. The intensity scaling is logarithmic. }
\label{fig:astralux_stamps}
\end{figure*}

The composite Figure~\ref{fig:astralux_stamps} shows six systems from the AstraLux observations, presenting a stellar pair of similar brightness, a pair of a bright and a faint component, a wide, a tight pair and the two triple systems.

\subsection{SPHERE}

We also obtained observations of the V1000 Tau system during the Science Verification of SPHERE \citep{2008SPIE.7014E..18B,2008SPIE.7014E..3LD,2010MNRAS.407...71V,2008SPIE.7014E..3EC}, the newly installed extreme adaptive optics facility at the VLT. 
These observations were acquired on 2014 Dec. 10$^\mathrm{th}$, when SPHERE was operated in the IRDIFS mode with a 155 milli-arcsecond (mas) diameter apodized Lyot coronagraph, offering simultaneous Integral Field Spectrograph (IFS) observations from 0.95--1.65~$\mu$m, and imaging in the Ks ($2.181\ \mathrm{\mu m}$) filter. We also obtained observations with the star offset from the coronagraphic mask  to be able to measure total fluxes. 
A PSF reference star (\object{TYC 1290-457-1}) was observed right after the IRDIFS observation of V1000 Tau, at similar airmass and atmospheric conditions as the science target.
The obtained data sets were reduced with the pre-release version 0.15.0-2 of the SPHERE pipeline. 
Apart from the usual steps of bias subtraction, flat-fielding, and wavelength calibration, the pipeline also includes a correction for geometric distortions. In the calculated positions of the stars, we also took into account the detector's deviation from true north orientation in the position angle ($-1.788^\circ \pm 0.008^\circ$, detailed in the SPHERE manual\footnote{\url{http://www.eso.org/sci/facilities/paranal/instruments/sphere/doc/VLT-MAN-SPH-14690-0430_v96.pdf}, Section~9.3.2, page 64}). Post-processing of the data was done using the FITSH software package \citep{2012MNRAS.421.1825P} to sum the data resulting from the two IRDIS channels. 

\subsection{NACO}

We also took advantage of the extensive ESO archive and searched for publicly available data on the given systems. We found 19, previously unpublished, adaptive optics assisted imaging observations of our targets captured by VLT/NACO \citep{2003SPIE.4841..944L,2003SPIE.4839..140R}. We reduced the previously unpublished  observations (that employed standard J, H, K and narrow band 1.64~$\mu$m, 2.12~$\mu$m, 2.17~$\mu$m filters) with the ESO provided NACO pipeline\footnote{\url{http://www.eso.org/sci/software/pipelines/}} (version 4.4.0), achieving $\approx 65-85$~mas FWHM resolution in the K-band. These observations are also included in our analysis.

\subsection{Auxiliary data from the literature}

A literature search for spatially resolved data of the systems in our sample resulted in including measurements from papers listed in Table~\ref{table:obs}. The most extensive literature data sets come from \citet{2001ApJ...556..265W}, \cite{2006ApJ...636..932M} and \citet{2006AnA...459..909C}. \cite{2006ApJ...636..932M} observed 65 T Tauri binaries using mid- and near-infrared adaptive optics instruments on the Keck 10~m telescopes between May 1998 and November 2002 and \citet{2006AnA...459..909C} observed T Tauri triple and quadruple systems employing VLT/NACO.
The observations of the objects are listed in Table~\ref{table:obs}.

\begin{longtab}
\setlength{\tabcolsep}{3pt}
\begin{longtable}{lllllllll}
\caption{\label{table:obs}Astrometric epochs of each pair in our sample. The magnitude difference between the two stars are also shown where we had photometric data. For a few epochs, we had no precise observation date available, in those cases we used a period in which the data could have been acquired and use the middle of that period in the calculations. }\\
\hline\hline
Object & Date & Telescope & Method$^1$ & Filter & Sep. ($''$) & PA ($^\circ$) & Magn. diff. & Ref. \\
\hline
\endfirsthead
\caption{Astrometric epochs continued.}\\
\hline\hline
Object & Date & Telescope & Method$^1$ & Filter & Sep. ($''$) & PA ($^\circ$) & Magn. diff. & Ref.\\
\hline
\endhead
\hline
\multicolumn{9}{l}{\multirow{2}{0.97\textwidth}{References:
1:~\cite{1944PASP...56..123J}, 2:~\cite{1989PhDT........11W}, 3:~\cite{1990A&A...230L...1H}, 4:~\cite{1991A&A...242..428M}, 5:~\cite{1991A&A...250..407L}, 6:~\cite{1992A&A...261..451B}, 7:~\cite{1992ApJ...384..212S}, 8:~\cite{1992ESOC...39...47Z}, 9:~\cite{1993AJ....106.2005G}, 10:~\cite{1993AnA...278..129L}, 11:~\cite{1994AnA...283..827T}, 12:~\cite{1994ApJ...427..961H}, 13:~\cite{1994ApJ...434..707G}, 14:~\cite{1995AJ....110..753G}, 15:~\cite{1997ApJ...478..766C}, 16:~\cite{1997ApJ...481..447K}, 17:~\cite{1998A&A...338..122W}, 18:~\cite{1999ApJ...515L..35K}, 19:~\cite{2001A&A...369..249W}, 20:~\cite{2001ApJ...556..265W}, 21:~\cite{2003AnA...400..559D}, 22:~\cite{2003ApJ...583..334H}, 23:~\cite{2003ApJ...588L.113M}, 24:~\cite{2006AnA...452..897C}, 25:~\cite{2006AnA...459..909C}, 26:~\cite{2006ApJ...636..932M}, 27:~\cite{2008AJ....135.2496C}, 28:~\cite{2008AJ....136.1980K}, 29:~\cite{2009ApJ...693L..86C}, 30:~\cite{2010ApJ...712..112D}, 31:~\cite{2012ARep...56..686B}, 32:~\cite{2013ApJ...773...40T}, 33:~\cite{2014AJ....147..157S}, 34:~\cite{2014ApJ...784...62A}, 35:~\cite{2014ApJ...788..101S}, 36:~\cite{2014MNRAS.439.4057F}, 37:~\cite{2015ApJ...799..155D}, 38:~\cite{2015ApJ...808L...3A}, 39:~\cite{2016AstL...42...29D}, 40:~this~work.
}}\\[9.2em]
\multicolumn{9}{l}{\multirow{2}{0.97\textwidth}{$^1$ Method: SI: speckle interferometry, LI: lucky imaging, I: imaging, AO: adaptive optics, XAO: extreme adaptive optics, 
R: radio observations, X: X-ray observations
}}\\[1.2em]
\multicolumn{9}{l}{\multirow{2}{0.97\textwidth}{$^2$ FV Tau B has two data points from 1999-11-17 which do not agree within the uncertainties, therefore we dismissed the measurement that is clearly an outlier.
}}\\[1.2em]
\multicolumn{9}{l}{\multirow{2}{0.97\textwidth}{$^3$ We obtained the values of UX Tau AB from the separations and positions angles of UX Tau A-Ba and UX Tau A-Bb as listed in \citet{2014AJ....147..157S} by linearly interpolating the positions using the flux ratio provided for UX Tau Ba-Bb therein. 
}}\\[1.2em]
\multicolumn{9}{l}{\multirow{2}{0.97\textwidth}{$^4$ We recalculated  the 1990-10-08 epoch of DK Tau, because their numbers strongly suggest a coordinate conversion error. We also recalculated the 2012-11-17 epoch of DK Tau, because the quoted separation does not agree with the quoted coordinates.
}}\\[1.2em]
\multicolumn{9}{l}{\multirow{2}{0.97\textwidth}{$^5$ We re-analysed these observations using the current NACO pipeline to be able to draw conclusions from a consistently analysed data set.
}}\\[1.2em]
\multicolumn{9}{l}{\multirow{2}{0.97\textwidth}{$^6$ These dates are the middle of the time interval 1990-12 -- 1992-10 provided by \citet{1993AnA...278..129L} for the time of their observations. 
}}\\[1.2em]
\multicolumn{9}{l}{\multirow{2}{0.97\textwidth}{$^7$ We opted to re-analyse the original VLT/NACO images using the current NACO pipeline, because the original paper does not feature a magnitude difference measurement, but the companion is actually clearly visible and measurable.
}}\\[2em]
\multicolumn{9}{l}{\multirow{2}{0.97\textwidth}{Other notes: We did not include the binary survey by \citet{1993AnA...278...81R}, because their measurements do not cite accurate dates  (only a six year long interval) and also does not cite astrometric errors. $\bullet$ We have omitted the observation of the UX Tau AC pair by \citet{1992RMxAA..24..109D} because they do not give any hint on the actual observation date. $\bullet$ We have omitted the observation of FV Tau B by \citet{1990ApJ...357..224C} because they employed lunar occultation and they were only able to determine the separation along the direction of the occultation. $\bullet$ We have omitted the observation of V710 Tau, HN Tau and HK Tau by \citet{1979ApJS...41..743C} because they do not list the uncertainties of their measurements.
}}\\[1.5em]
\endlastfoot
\hline
\endfoot
       LkCa 3 B & 1990-12-06 & CAHA 3.5m & SI & K & $0.470 \pm 0.040  $ & $  78.0 \pm 1.0  $ & $ 0.753 \pm 0.055 $ & 19 \\
                & 1991-10-18 & Palomar/Hale 5.1m & SI & K & $0.491 \pm 0.009  $ & $  77.0 \pm 1.0  $ & $ 0.049 \pm 0.006 $ & 9 \\
                & 1993-10-05 & CAHA 3.5m & SI & J & $0.494 \pm 0.013  $ & $  74.5 \pm 0.7  $ & $ 0.519 \pm 0.036 $ & 19 \\
                & 1996-09-29 & CAHA 3.5m & SI & H & $0.485 \pm 0.004  $ & $  74.2 \pm 0.3  $ & $ 0.170 \pm 0.014 $ & 19 \\
                & 1997-11-19 & CAHA 3.5m & SI & K & $0.485 \pm 0.004  $ & $  73.4 \pm 0.1  $ & $ 0.305 \pm 0.011 $ & 19 \\
                & 1997-12-05 & NASA ITF/NSFCAM & SI & K & $0.479 \pm 0.010  $ & $  72.8 \pm 1.0  $ & $ 0.225 \pm 0.044 $ & 20 \\
                & 1997-12-09  & NASA ITF/NSFCAM & SI & L & $0.478 \pm 0.009  $ & $  73.0 \pm 1.4  $ & $ 0.120 \pm 0.044 $ & 20 \\
                & 2001-12-01 & Keck & I & K & $0.470 \pm 0.010  $ & $  72.0 \pm 1.0  $ & $ 0.180 \pm 0.028 $ & 26 \\
                & 2001-12-01  & Keck & I & L & $0.470 \pm 0.010  $ & $  72.0 \pm 1.0  $ & $ 0.170 \pm 0.010 $ & 26 \\
                & 2002-11-13 & Keck & I & SiC & $0.480 \pm 0.020  $ & $  76.0 \pm 2.0  $ & $ 0.373 \pm 0.047 $ & 26 \\
                & 2006-11-11 & CAHA 2.2m & LI & RG830 & $0.477 \pm 0.020  $ & $  68.7 \pm 1.9  $ & $ 0.575 \pm 0.021 $ & 40 \\
                & 2008-01-17 & Keck & AO & K & $0.484 \pm 0.001  $ & $  69.1 \pm 0.1  $ & $ 0.067 \pm 0.009 $ & 32 \\[0.5em]
       DD Tau B & 1990-10-23 & CFHT & I & H$-\alpha$, cont & $0.550 \pm 0.030  $ & $ 184.0 \pm 2.0  $ & $ 0.450 \pm 0.101 $ & 6 \\
                & 1990-11-06 & CFHT & I & R & $0.560 \pm 0.020  $ & $ 184.0 \pm 2.0  $ & $ 0.100 \pm 0.101 $ & 6 \\
                & 1990-11-08 & CFHT & I & 6840/90 & $0.560 \pm 0.040  $ & $ 185.0 \pm 3.0  $ & $ 0.000 \pm 0.201 $ & 6 \\
                & 1990-12-05 & CFHT & SI & L & $0.560 \pm 0.010  $ & $ 186.2 \pm 1.5  $ & $ 0.649 \pm 0.040 $ & 6, 11 \\
                & 1990-12-15 & CAHA 3.5m & SI & K & $0.570 \pm 0.030  $ & $ 188.0 \pm 2.0  $ & $ 0.485 \pm 0.017 $ & 8, 10 \\
                & 1991-10-04 & Palomar/Hale 5.1m & SI & K & $0.560 \pm 0.010  $ & $ 186.0 \pm 1.0  $ & $ 0.708 \pm 0.012 $ & 9 \\
                & 1991-10-11 & CFHT & I & V & $0.560 \pm 0.020  $ & $ 182.0 \pm 3.0  $ & $ 0.000 \pm 0.151 $ & 6 \\
                & 1991-10-11  & CFHT & I & I & $0.550 \pm 0.020  $ & $ 184.0 \pm 2.0  $ & $ 0.000 \pm 0.101 $ & 6 \\
                & 1996-12-05 & NASA ITF/NSFCAM & SI & K & $0.555 \pm 0.010  $ & $ 182.3 \pm 1.0  $ & $ 0.173 \pm 0.055 $ & 20 \\
                & 1996-12-06  & NASA ITF/NSFCAM & SI & L & $0.557 \pm 0.011  $ & $ 182.7 \pm 1.1  $ & $ 0.589 \pm 0.083 $ & 20 \\
                & 1997-12-05 & NASA ITF/NSFCAM & SI & K & $0.555 \pm 0.010  $ & $ 182.1 \pm 1.0  $ & $ 0.402 \pm 0.035 $ & 20 \\
                & 1999-09-06 & HST & I & Clear & $0.568 \pm 0.014  $ & $ 182.1 \pm 1.0  $ & -- & 22 \\
                & 1999-11-17 & Keck & I & N & $0.540 \pm 0.010  $ & $ 179.0 \pm 2.0  $ & $ 0.608 \pm 0.050 $ & 26 \\
                & 2006-11-11 & CAHA 2.2m & LI & Johnson I & $0.568 \pm 0.034  $ & $ 179.0 \pm 3.3  $ & $ 1.123 \pm 0.055 $ & 40 \\[0.5em]
       LkCa 7 B & 1991-09-15 & CAHA 3.5m & SI & K & $1.050 \pm 0.010  $ & $  25.0 \pm 2.0  $ & $ 0.630 \pm 0.039 $ & 8, 10 \\
                & 1997-12-09 & NASA ITF/NSFCAM & SI & L & $1.021 \pm 0.019  $ & $  24.3 \pm 1.0  $ & $ 0.594 \pm 0.059 $ & 20 \\
                & 1999-01-26 & HST & I & Clear & $1.035 \pm 0.014  $ & $  25.2 \pm 0.6  $ & -- & 22 \\
                & 2001-12-01 & Keck & I & K & $0.980 \pm 0.020  $ & $  25.0 \pm 1.0  $ & $ 0.669 \pm 0.030 $ & 26 \\
                & 2001-12-01  & Keck & I & L & $0.990 \pm 0.020  $ & $  25.0 \pm 1.0  $ & $ 0.677 \pm 0.012 $ & 26 \\
                & 2006-11-11 & CAHA 2.2m & LI & Johnson I & $1.006 \pm 0.031  $ & $  23.9 \pm 1.8  $ & $ 0.881 \pm 0.054 $ & 40 \\[0.5em]
       FV Tau B & 1989-10-17 & CAHA 3.5m & O & K & $0.735 \pm 0.015  $ & $ 271.5 \pm 1.5  $ & $ 0.182 \pm 0.010 $ & 5 \\
                & 1990-07-18 & IRTF & O & K & $0.720 \pm 0.010  $ & $ 270.0 \pm 5.0  $ & $ 0.200 \pm 0.011 $ & 7 \\
                & 1991-10-20 & Palomar/Hale 5.1m & SI & K & $0.730 \pm 0.010  $ & $ 272.0 \pm 1.0  $ & $ 0.483 \pm 0.028 $ & 9 \\
                & 1996-12-05 & NASA ITF/NSFCAM & SI & K & $0.704 \pm 0.013  $ & $ 271.6 \pm 1.0  $ & $ 0.429 \pm 0.008 $ & 20 \\
                & 1996-12-05  & NASA ITF/NSFCAM & SI & L & $0.712 \pm 0.014  $ & $ 271.9 \pm 1.0  $ & $ 0.408 \pm 0.031 $ & 20 \\
                & 1997-03-06 & HST & I & F814W & $0.718 \pm 0.002  $ & $ 272.5 \pm 0.2  $ & $ 0.891 \pm 0.006 $ & 20 \\
                & 1999-11-17 & Keck & I & N & $0.710 \pm 0.020 ^2 $ & $ 272.1 \pm 0.8 ^2 $ & $ 0.856 \pm 0.099 $ & 26 \\
                & 2000-12-02 & HST & I & Clear & $0.711 \pm 0.014  $ & $ 273.0 \pm 0.8  $ & -- & 22 \\
                & 2002-11-13 & VLT/NACO & I & NB\_1.64 & $0.703 \pm 0.004  $ & $ 273.3 \pm 0.2  $ & $ 1.818 \pm 0.003 $ & 25, 40 \\
                & 2002-11-13  & VLT/NACO & I & NB\_2.12 & $0.707 \pm 0.004  $ & $ 273.1 \pm 0.2  $ & $ 1.299 \pm 0.005 $ & 25, 40 \\
                & 2002-11-13   & VLT/NACO & I & NB\_2.17 & $0.710 \pm 0.004  $ & $ 273.2 \pm 0.2  $ & $ 1.276 \pm 0.006 $ & 25, 40 \\
                & 2006-11-11 & CAHA 2.2m & LI & Johnson I & $0.684 \pm 0.041  $ & $ 272.2 \pm 1.6  $ & $ 2.677 \pm 0.058 $ & 40 \\
                & 2012-09-25 & Gemini North & AO & K & $0.693 \pm 0.001  $ & $ 274.5 \pm 0.1  $ & $ 1.010 \pm 0.011 $ & 37 \\
                & 2012-11-17 & ALMA & R & 1.3mm & $0.696 \pm 0.046  $ & $ 274.4 \pm 3.0  $ & $ 0.043 \pm 0.044 $ & 34 \\
                & 2012-11-17  & ALMA & R & 850 $\mu$m & $0.696 \pm 0.046  $ & $ 274.4 \pm 3.0  $ & $ 0.140 \pm 0.065 $ & 34 \\[0.5em]
     FV Tau/c B & 1990-07-18 & IRTF & O & K & $0.743 \pm 0.140  $ & $ 293.0 \pm 3.0  $ & $ 1.900 \pm 0.096 $ & 7 \\
                & 1996-12-05 & NASA ITF/NSFCAM & SI & K & $0.701 \pm 0.014  $ & $ 293.3 \pm 1.0  $ & $ 2.211 \pm 0.029 $ & 20 \\
                & 1996-12-06  & NASA ITF/NSFCAM & SI & L & $0.703 \pm 0.014  $ & $ 293.8 \pm 1.0  $ & $ 1.160 \pm 0.056 $ & 20 \\
                & 1997-03-06 & HST & I & F814W & $0.713 \pm 0.002  $ & $ 294.0 \pm 0.1  $ & $ 3.696 \pm 0.116 $ & 20 \\
                & 1998-12-06 & HST & I & Clear & $0.701 \pm 0.014  $ & $ 293.6 \pm 0.8  $ & -- & 22 \\
                & 1999-11-17 & Keck & I & N & $0.670 \pm 0.020  $ & $ 292.0 \pm 1.0  $ & $ 0.327 \pm 0.133 $ & 26 \\
                & 2002-11-12 & Keck & I & SiC & $0.680 \pm 0.020  $ & $ 295.0 \pm 1.0  $ & $ 0.312 \pm 0.058 $ & 26 \\
                & 2002-11-13 & VLT/NACO & I & NB\_1.64 & $0.688 \pm 0.004  $ & $ 294.5 \pm 0.3  $ & $ 1.924 \pm 0.003 $ & 25, 40 \\
                & 2002-11-13  & VLT/NACO & I & NB\_2.12 & $0.688 \pm 0.004  $ & $ 294.5 \pm 0.3  $ & $ 1.632 \pm 0.006 $ & 25, 40 \\
                & 2002-11-13   & VLT/NACO & I & NB\_2.17 & $0.689 \pm 0.004  $ & $ 294.6 \pm 0.3  $ & $ 1.746 \pm 0.006 $ & 25, 40 \\
                & 2006-11-11 & CAHA 2.2m & LI & Johnson I & $0.630 \pm 0.044  $ & $ 294.4 \pm 3.2  $ & $ 3.111 \pm 0.060 $ & 40 \\[0.5em]
      UX Tau AC & 1988-01-15 & CTIO & I & K & $2.700 \pm 0.100  $ & $ 181.0 \pm 2.0  $ & $ 2.910 \pm 0.021 $ & 4 \\
                & 1991-10-27 & Kitt Peak/0.9m & I & R & $2.600 \pm 0.053  $ & $ 182.0 \pm 1.0  $ & $ 4.430 \pm 0.191 $ & 12 \\
                & 1997-12-06 & NASA ITF/NSFCAM & I & K & $2.632 \pm 0.027  $ & $ 180.2 \pm 1.0  $ & $ 3.070 \pm 0.103 $ & 20 \\
                & 1997-12-06  & NASA ITF/NSFCAM & I & L & $2.634 \pm 0.033  $ & $ 180.2 \pm 1.0  $ & $ 3.451 \pm 0.091 $ & 20 \\
                & 2002-10-22  & VLT/NACO & I & NB\_1.64 & $2.684 \pm 0.005  $ & $ 181.5 \pm 0.1  $ & $ 3.392 \pm 0.002 $ & 25, 40 \\
                & 2002-10-22   & VLT/NACO & I & NB\_2.12 & $2.684 \pm 0.005  $ & $ 181.6 \pm 0.1  $ & $ 2.970 \pm 0.001 $ & 25, 40 \\
                & 2002-10-22    & VLT/NACO & I & NB\_2.17 & $2.682 \pm 0.005  $ & $ 181.6 \pm 0.1  $ & $ 3.135 \pm 0.001 $ & 25, 40 \\
                & 2002-11-12 & Keck & I & SiC & $2.500 \pm 0.300  $ & $ 181.0 \pm 5.0  $ & $ 4.440 \pm 0.131 $ & 26 \\
                & 2006-11-12 & CAHA 2.2m & LI & RG830 & $2.705 \pm 0.047  $ & $ 180.5 \pm 0.5  $ & $ 3.478 \pm 0.055 $ & 40 \\
                & 2009-10-26 & Keck II & AO & K & $2.711 \pm 0.004  $ & $ 181.6 \pm 0.1  $ & $ 3.212 \pm 0.107 $ & 33 \\
                & 2011-10-12 & Keck II & AO & K & $2.716 \pm 0.002  $ & $ 181.6 \pm 0.1  $ & $ 3.122 \pm 0.062 $ & 33 \\
                & 2013-01-28 & Keck II & AO & K & $2.720 \pm 0.003  $ & $ 181.6 \pm 0.1  $ & $ 2.845 \pm 0.182 $ & 33 \\[0.5em]
      UX Tau AB & 1944-04-01 & McDonald/2.1m & I & Z & $5.740 \pm 0.109  $ & $ 270.1 \pm 0.2  $ & -- & 1 \\
                & 1988-01-15 & CTIO & I & K & $5.900 \pm 0.100  $ & $ 269.0 \pm 2.0  $ & $ 1.350 \pm 0.021 $ & 4 \\
                & 1991-10-27 & Kitt Peak/0.9m & I & R & $5.800 \pm 0.116  $ & $ 270.0 \pm 1.0  $ & $ 1.840 \pm 0.016 $ & 12 \\
                & 1997-12-06 & NASA ITF/NSFCAM & I & K & $5.860 \pm 0.110  $ & $ 269.4 \pm 1.0  $ & $ 1.480 \pm 0.087 $ & 20 \\
                & 1997-12-06  & NASA ITF/NSFCAM & I & L & $5.860 \pm 0.110  $ & $ 269.3 \pm 1.0  $ & $ 2.149 \pm 0.029 $ & 20 \\
                & 2002-10-22  & VLT/NACO & I & NB\_1.64 & $5.846 \pm 0.009  $ & $ 269.7 \pm 0.1  $ & $ 2.216 \pm 0.004 $ & 25, 40 \\
                & 2002-10-22   & VLT/NACO & I & NB\_2.12 & $5.839 \pm 0.009  $ & $ 269.7 \pm 0.1  $ & $ 2.188 \pm 0.002 $ & 25, 40 \\
                & 2002-10-22    & VLT/NACO & I & NB\_2.17 & $5.841 \pm 0.009  $ & $ 269.7 \pm 0.1  $ & $ 2.097 \pm 0.003 $ & 25, 40 \\
                & 2002-11-12 & Keck & I & SiC & $5.600 \pm 0.100  $ & $ 268.0 \pm 1.0  $ & $ 2.689 \pm 0.165 $ & 26 \\
                & 2006-11-12 & CAHA 2.2m & LI & RG830 & $5.874 \pm 0.053  $ & $ 268.6 \pm 0.5  $ & $ 1.422 \pm 0.054 $ & 40 \\
                & 2009-10-26 & Keck II & AO & K & $5.921 \pm 0.003 ^3 $ & $ 270.1 \pm 0.1 ^3 $ & $ 2.269 \pm 0.054 $ & 33 \\
                & 2011-10-12 & Keck II & AO & K & $5.908 \pm 0.002 ^3 $ & $ 269.9 \pm 0.1 ^3 $ & $ 2.181 \pm 0.081 $ & 33 \\
                & 2013-01-28 & Keck II & AO & K & $5.906 \pm 0.007 ^3 $ & $ 269.9 \pm 0.1 ^3 $ & $ 2.354 \pm 0.313 $ & 33 \\[0.5em]
       FX Tau B & 1990-12-15 & CAHA 3.5m & SI & K & $0.910 \pm 0.010  $ & $ 292.0 \pm 3.0  $ & $ 0.649 \pm 0.020 $ & 8, 10 \\
                & 1991-10-18 & Palomar/Hale 5.1m & SI & K & $0.900 \pm 0.020  $ & $ 291.0 \pm 1.0  $ & $ 0.733 \pm 0.001 $ & 9 \\
                & 1997-12-07 & NASA ITF/NSFCAM & SI & K & $0.890 \pm 0.017  $ & $ 289.0 \pm 1.0  $ & $ 0.857 \pm 0.008 $ & 20 \\
                & 1997-12-07  & NASA ITF/NSFCAM & SI & L & $0.885 \pm 0.017  $ & $ 289.0 \pm 1.0  $ & $ 1.182 \pm 0.013 $ & 20 \\
                & 1999-11-18 & Keck & I & N & $0.870 \pm 0.020  $ & $ 289.4 \pm 0.4  $ & $ 1.098 \pm 0.044 $ & 26 \\
                & 1999-11-18  & Keck & I & IHW18 & $0.840 \pm 0.020  $ & $ 290.5 \pm 0.5  $ & $ 1.114 \pm 0.129 $ & 26 \\
                & 2001-12-01 & Keck & I & L & $0.850 \pm 0.020  $ & $ 289.0 \pm 1.0  $ & $ 1.050 \pm 0.034 $ & 26 \\
                & 2006-11-11 & CAHA 2.2m & LI & Johnson I & $0.864 \pm 0.075  $ & $ 286.6 \pm 4.6  $ & $ 1.625 \pm 0.059 $ & 40 \\[0.5em]
       DK Tau B & 1988-10-20 & Lick & SI & K & $2.800 \pm 0.300  $ & $ 115.0 \pm 7.0  $ & $ 1.505 \pm 0.435 $ & 2 \\
                & 1990-10-08 & IRTF & O & K & $2.330 \pm 0.050  $ & $ 120.0 \pm 2.0  $ & $ 1.300 \pm 0.101 $ & 7 \\
                & 1997-12-06 & NASA ITF/NSFCAM & I & K & $2.304 \pm 0.045  $ & $ 117.6 \pm 1.2  $ & $ 1.561 \pm 0.078 $ & 20 \\
                & 1997-12-06  & NASA ITF/NSFCAM & I & L & $2.337 \pm 0.056  $ & $ 117.6 \pm 1.3  $ & $ 1.511 \pm 0.036 $ & 20 \\
                & 2002-11-13 & Keck & I & SiC & $2.270 \pm 0.060  $ & $ 117.0 \pm 1.0  $ & $ 2.327 \pm 0.006 $ & 26 \\
                & 2002-11-13   & VLT/NACO & I & NB\_1.64 & $2.354 \pm 0.005  $ & $ 119.1 \pm 0.2  $ & $ 1.150 \pm 0.008 $ & 25, 40 \\
                & 2002-11-13    & VLT/NACO & I & NB\_2.12 & $2.350 \pm 0.005  $ & $ 119.1 \pm 0.1  $ & $ 1.373 \pm 0.004 $ & 25, 40 \\
                & 2002-11-13     & VLT/NACO & I & NB\_2.17 & $2.350 \pm 0.005  $ & $ 119.1 \pm 0.1  $ & $ 1.386 \pm 0.004 $ & 25, 40 \\
                & 2006-11-11 & CAHA 2.2m & LI & Johnson I & $2.358 \pm 0.076  $ & $ 118.2 \pm 1.9  $ & $ 1.422 \pm 0.054 $ & 40 \\
                & 2012-11-17 & ALMA & R & 1.3mm & $2.382 \pm 0.047 ^4 $ & $ 118.7 \pm 0.6 ^4 $ & $ 2.555 \pm 0.072 $ & 34 \\
                & 2012-11-17  & ALMA & R & 850 $\mu$m & $2.382 \pm 0.047 ^4 $ & $ 118.7 \pm 0.6 ^4 $ & $ 2.659 \pm 0.085 $ & 34 \\[0.5em]
       XZ Tau B & 1989-10-17 & CAHA 3.5m & SI & K & $0.300 \pm 0.020  $ & $ 154.0 \pm 3.0  $ & $ 1.140 \pm 0.063 $ & 3 \\
                & 1991-10-20 & Palomar/Hale 5.1m & SI & K & $0.310 \pm 0.010  $ & $ 151.0 \pm 2.0  $ & $ 0.731 \pm 0.028 $ & 14 \\
                & 1994-01-27 & CAHA 3.5m & SI & J & $0.307 \pm 0.004  $ & $ 146.1 \pm 0.6  $ & $ 0.447 \pm 0.022 $ & 19 \\
                & 1994-01-28 & CAHA 3.5m & SI & K & $0.306 \pm 0.005  $ & $ 146.8 \pm 1.0  $ & $ 0.968 \pm 0.027 $ & 19 \\
                & 1994-12-19 & Palomar/Hale 5.1m & SI & K & $0.296 \pm 0.002  $ & $ 147.0 \pm 0.4  $ & -- & 14 \\
                & 1995-01-05 & HST & I & R & $0.306 \pm 0.003  $ & $ 147.2 \pm 0.5  $ & $ 1.770 \pm 0.089 $ & 16, 28 \\
                & 1995-11-15 & CFHT 3.6m & I & K & $0.290 \pm 0.030  $ & $ 147.0 \pm 5.0  $ & $ 0.890 \pm 0.121 $ & 15 \\
                & 1996-09-29 & CAHA 3.5m & SI & K & $0.299 \pm 0.004  $ & $ 145.4 \pm 0.3  $ & -- & 19 \\
                & 1996-11-30 & CAHA 3.5m & SI & J & $0.309 \pm 0.010  $ & $ 145.7 \pm 0.7  $ & $ 1.373 \pm 0.096 $ & 19 \\
                & 1996-12-05 & NASA ITF/NSFCAM & SI & K & $0.300 \pm 0.006  $ & $ 144.5 \pm 1.0  $ & $ 0.556 \pm 0.040 $ & 20 \\
                & 1996-12-05  & NASA ITF/NSFCAM & SI & L & $0.299 \pm 0.029  $ & $ 142.1 \pm 1.1  $ & $ 1.307 \pm 0.174 $ & 20 \\
                & 1997-03-08 & HST & I & F814W & $0.301 \pm 0.002  $ & $ 146.4 \pm 0.7  $ & $ 2.113 \pm 0.012 $ & 20 \\
                & 1997-11-19 & CAHA 3.5m & SI & K & $0.302 \pm 0.004  $ & $ 140.9 \pm 0.5  $ & $ 1.251 \pm 0.025 $ & 19 \\
                & 1998-03-25 & HST & I & R & $0.302 \pm 0.009  $ & $ 145.5 \pm 0.3  $ & $ 2.590 \pm 0.130 $ & 18 \\
                & 1998-12-01 & HST & I & R & $0.298 \pm 0.003  $ & $ 145.5 \pm 0.5  $ & $ 2.590 \pm 0.130 $ & 28 \\
                & 1999-02-03 & HST & I & R & $0.301 \pm 0.003  $ & $ 144.4 \pm 0.5  $ & $ 1.500 \pm 0.076 $ & 28 \\
                & 2000-02-06 & HST & I & R & $0.299 \pm 0.003  $ & $ 142.6 \pm 0.5  $ & $ 1.020 \pm 0.052 $ & 28 \\
                & 2000-12-02 & HST & I & Clear & $0.299 \pm 0.014  $ & $ 142.6 \pm 2.0  $ & -- & 22 \\
                & 2001-02-10 & HST & I & R & $0.297 \pm 0.003  $ & $ 141.6 \pm 0.5  $ & $ 0.500 \pm 0.026 $ & 28 \\
                & 2002-02-12 & HST & I & R & $0.291 \pm 0.003  $ & $ 141.6 \pm 0.5  $ & $ 2.230 \pm 0.112 $ & 28 \\
                & 2003-12-13 & VLT/NACO & I & Ks & $0.292 \pm 0.025  $ & $ 139.3 \pm 4.8  $ & $ 1.445 \pm 0.001 $ & 25, 40 \\
                & 2003-12-13  & VLT/NACO & I & J & $0.289 \pm 0.086  $ & $ 139.7 \pm 16.6  $ & $ 0.911 \pm 0.002 $ & 25, 40 \\
                & 2004-01-20 & HST & I & R & $0.294 \pm 0.003  $ & $ 142.3 \pm 0.5  $ & $ 2.710 \pm 0.136 $ & 28 \\
                & 2004-01-22 & VLA & R & 7mm & $0.301 \pm 0.006  $ & $ 138.6 \pm 0.5  $ & $ 1.505 \pm 0.747 $ & 29 \\
                & 2006-11-11 & CAHA 2.2m & LI & Johnson I & $0.256 \pm 0.100  $ & $ 135.4 \pm 20.6  $ & $ 0.935 \pm 0.054 $ & 40 \\
                & 2006-11-20 & VLT/NACO & I & Ks & $0.289 \pm 0.005  $ & $ 136.9 \pm 0.2  $ & $ 0.618 \pm 0.002 $ & 25, 40 \\
                & 2012-10-11 & VLA & R & 7mm & $0.282 \pm 0.004  $ & $ 129.7 \pm 0.1  $ & $ 0.469 \pm 0.192 $ & 36 \\
                & 2014-02-14 & BTA-6 & SI & I & $0.270 \pm 0.003  $ & $ 131.5 \pm 0.7  $ & -- & 39 \\
                & 2014-04-11 & BTA-6 & SI & I & $0.268 \pm 0.004  $ & $ 131.4 \pm 0.9  $ & $ 0.370 \pm 0.351 $ & 39 \\
                & 2014-10-14 & ALMA & R & 2.9mm & $0.273 \pm 0.001  $ & $ 128.7 \pm 0.5  $ & $ 0.462 \pm 0.040 $ & 38 \\[0.5em]
       HK Tau B & 1988-01-15 & CTIO & I & K & $2.400 \pm 0.100  $ & $ 175.0 \pm 2.0  $ & $ 3.070 \pm 0.021 $ & 4 \\
                & 1997-12-06 & NASA ITF/NSFCAM & I & K & $2.342 \pm 0.061  $ & $ 170.4 \pm 1.1  $ & $ 3.321 \pm 0.057 $ & 20 \\
                & 1997-12-06  & NASA ITF/NSFCAM & I & L & $2.284 \pm 0.053  $ & $ 168.8 \pm 2.4  $ & $ 3.678 \pm 0.114 $ & 20 \\
                & 1998-12-17 & IRAM & R & 1.4mm & $2.440 \pm 0.050  $ & $ 171.7 \pm 1.1  $ & $ 0.582 \pm 0.084 $ & 21 \\
                & 2001-12-01 & Keck & I & L & $2.390 \pm 0.100  $ & $ 168.1 \pm 2.7  $ & $ 3.516 \pm 0.103 $ & 26 \\
                & 2002-11-13 & Keck & I & SiC & $2.228 \pm 0.056  $ & $ 170.3 \pm 1.2  $ & $ 3.698 \pm 0.172 $ & 23 \\
                & 2002-11-19 & VLT/NACO & I & Ks & $2.413 \pm 0.060 ^5 $ & $ 169.8 \pm 1.2 ^5 $ & $ 2.442 \pm 0.001 $ & 25, 40 \\
                & 2002-11-19  & VLT/NACO & I & NB\_2.12 & $2.421 \pm 0.057 ^5 $ & $ 170.0 \pm 1.1 ^5 $ & $ 4.191 \pm 0.001 $ & 25, 40 \\
                & 2004-02-11 & UKIRT/UFTI & I & L & $2.300 \pm 0.010  $ & $ 171.4 \pm 0.1  $ & $ 3.500 \pm 0.051 $ & 27 \\
                & 2006-11-11 & CAHA 2.2m & LI & Johnson I & $2.305 \pm 0.093  $ & $ 169.3 \pm 1.6  $ & $ 3.283 \pm 0.057 $ & 40 \\
                & 2012-11-17 & ALMA & R & 1.3mm & $2.349 \pm 0.036  $ & $ 170.1 \pm 0.5  $ & $ 0.811 \pm 0.018 $ & 34 \\
                & 2012-11-17  & ALMA & R & 850 $\mu$m & $2.349 \pm 0.036  $ & $ 170.1 \pm 0.5  $ & $ 0.287 \pm 0.040 $ & 34 \\[0.5em]
     V710 Tau B & 1990-11-26 & San Pedro Martir/0.9m & I & R & $3.300 \pm 0.066  $ & $ 179.0 \pm 1.0  $ & $ 0.120 \pm 0.016 $ & 12 \\
                & 1991-11-01$^6$ & CAHA 3.5m & SI & K & $3.240 \pm 0.100  $ & $ 177.0 \pm 1.0  $ & $ 0.202 \pm 0.014 $ & 10 \\
                & 1997-12-06 & NASA ITF/NSFCAM & I & K & $3.168 \pm 0.062  $ & $ 176.2 \pm 1.1  $ & $ 0.270 \pm 0.042 $ & 20 \\
                & 1997-12-06  & NASA ITF/NSFCAM & I & L & $3.178 \pm 0.064  $ & $ 176.5 \pm 1.2  $ & $ 0.140 \pm 0.042 $ & 20 \\
                & 2002-10-22 & VLT/NACO & I & NB\_1.64 & $3.236 \pm 0.005  $ & $ 177.8 \pm 0.1  $ & $ 0.367 \pm 0.051 $ & 25, 40 \\
                & 2002-10-22  & VLT/NACO & I & NB\_2.12 & $3.229 \pm 0.004  $ & $ 177.8 \pm 0.1  $ & $ 0.007 \pm 0.072 $ & 25, 40 \\
                & 2002-10-22   & VLT/NACO & I & NB\_2.17 & $3.231 \pm 0.004  $ & $ 177.7 \pm 0.1  $ & $ 0.007 \pm 0.161 $ & 25, 40 \\
                & 2002-11-12 & Keck & I & SiC & $3.060 \pm 0.080  $ & $ 178.0 \pm 1.0  $ & $ 2.762 \pm 0.010 $ & 26 \\
                & 2006-11-12 & CAHA 2.2m & LI & RG830 & $3.226 \pm 0.042  $ & $ 176.7 \pm 0.5  $ & $ 0.175 \pm 0.054 $ & 40 \\[0.5em]
      UZ Tau AB & 1990-10-08 & IRTF & O & K & $0.340 \pm 0.060  $ & $   0.0 \pm 8.0  $ & $ 1.100 \pm 0.056 $ & 7 \\
                & 1990-11-10 & Palomar/Hale 5.1m & SI & K & $0.359 \pm 0.002  $ & $ 357.4 \pm 0.3  $ & -- & 14 \\
                & 1994-01-26 & CAHA 3.5m & SI & J & $0.358 \pm 0.010  $ & $   1.3 \pm 1.0  $ & $ 0.298 \pm 0.101 $ & 19 \\
                & 1994-07-24 & HST & I & F814W & $0.368 \pm 0.001  $ & $   0.1 \pm 0.2  $ & $ 0.292 \pm 0.025 $ & 20 \\
                & 1994-12-19 & Palomar/Hale 5.1m & SI & K & $0.360 \pm 0.003  $ & $   1.3 \pm 0.4  $ & -- & 14 \\
                & 1996-09-29 & CAHA 3.5m & SI & K & $0.366 \pm 0.004  $ & $   1.3 \pm 0.2  $ & $ 0.451 \pm 0.033 $ & 19 \\
                & 1996-12-06 & NASA ITF/NSFCAM & SI & K & $0.360 \pm 0.007  $ & $   0.8 \pm 1.0  $ & $ 0.544 \pm 0.066 $ & 20 \\
                & 1996-12-06  & NASA ITF/NSFCAM & SI & L & $0.369 \pm 0.007  $ & $   3.4 \pm 1.1  $ & $ 0.800 \pm 0.198 $ & 20 \\
                & 1999-11-03 & HST & I & Clear & $0.369 \pm 0.014  $ & $   3.6 \pm 1.6  $ & -- & 22 \\
                & 1999-11-16 & Keck & I & N & $0.370 \pm 0.010  $ & $   3.7 \pm 0.6  $ & $ 0.170 \pm 0.065 $ & 26 \\
                & 1999-11-17 & Keck & I & IHW18 & $0.370 \pm 0.010  $ & $   2.1 \pm 0.9  $ & $ 0.310 \pm 0.041 $ & 26 \\
                & 2001-12-01 & Keck & I & L & $0.350 \pm 0.010  $ & $   4.0 \pm 1.0  $ & $ 0.544 \pm 0.040 $ & 26 \\
                & 2002-11-14  & VLT/NACO & I & NB\_1.64 & $0.369 \pm 0.004  $ & $   6.2 \pm 0.6  $ & $ 0.662 \pm 0.089 $ & 25, 40 \\
                & 2002-11-14   & VLT/NACO & I & NB\_2.12 & $0.368 \pm 0.004  $ & $   5.8 \pm 0.5  $ & $ 0.868 \pm 0.057 $ & 25, 40 \\
                & 2002-11-14    & VLT/NACO & I & NB\_2.17 & $0.368 \pm 0.004  $ & $   5.4 \pm 0.5  $ & $ 0.668 \pm 0.051 $ & 25, 40 \\
                & 2006-11-11 & CAHA 2.2m & LI & Johnson I & $0.355 \pm 0.099  $ & $   5.6 \pm 10.2  $ & $ 0.540 \pm 0.054 $ & 40 \\
                & 2006-12-26 & VLT/NACO & I & Ks & $0.369 \pm 0.001  $ & $   8.1 \pm 0.2  $ & $ 0.588 \pm 0.003 $ & 40 \\
                & 2012-09-28 & Gemini North & AO & K & $0.375 \pm 0.001  $ & $  11.5 \pm 0.1  $ & $ 0.370 \pm 0.051 $ & 37 \\[0.5em]
      UZ Tau AC & 1944-04-01 & McDonald/2.1m & I & Z & $3.680 \pm 0.046  $ & $ 271.5 \pm 0.6  $ & $ 0.275 \pm 0.083 $ & 1 \\
                & 1988-01-15 & CTIO & I & K & $3.600 \pm 0.200  $ & $ 274.0 \pm 2.0  $ & $ 0.720 \pm 0.021 $ & 4 \\
                & 1990-10-08 & IRTF & O & K & $3.780 \pm 0.070  $ & $ 273.0 \pm 1.0  $ & $ 1.100 \pm 0.056 $ & 7 \\
                & 1990-11-26 & San Pedro Martir/0.9m & I & R & $3.500 \pm 0.070  $ & $ 275.0 \pm 1.0  $ & $ 0.390 \pm 0.016 $ & 12 \\
                & 1992-12-10 & UKIRT & I & 9.5$\mu$m & $3.600 \pm 0.100  $ & $ 275.0 \pm 2.0  $ & $ 1.659 \pm 0.194 $ & 13 \\
                & 1994-07-24 & HST & I & F814W & $3.539 \pm 0.003  $ & $ 273.1 \pm 0.1  $ & $ 0.276 \pm 0.186 $ & 20 \\
                & 1999-11-03 & HST & I & Clear & $3.551 \pm 0.014  $ & $ 273.2 \pm 0.2  $ & -- & 22 \\
                & 1999-11-16 & Keck & I & N & $3.520 \pm 0.060  $ & $ 272.9 \pm 0.4  $ & $ 2.260 \pm 0.120 $ & 26 \\
                & 1999-11-17 & Keck & I & IHW18 & $3.440 \pm 0.060  $ & $ 273.0 \pm 0.4  $ & $ 2.360 \pm 0.149 $ & 26 \\
                & 2001-12-01 & Keck & I & K & $3.500 \pm 0.100  $ & $ 275.0 \pm 1.0  $ & $ 0.644 \pm 0.048 $ & 26 \\
                & 2001-12-01  & Keck & I & L & $3.500 \pm 0.100  $ & $ 276.0 \pm 1.0  $ & $ 0.999 \pm 0.035 $ & 26 \\
                & 2002-11-14  & VLT/NACO & I & NB\_1.64 & $3.558 \pm 0.006  $ & $ 273.4 \pm 0.1  $ & $ 1.037 \pm 0.024 $ & 25, 40 \\
                & 2002-11-14   & VLT/NACO & I & NB\_2.12 & $3.550 \pm 0.006  $ & $ 273.5 \pm 0.1  $ & $ 1.477 \pm 0.009 $ & 25, 40 \\
                & 2002-11-14    & VLT/NACO & I & NB\_2.17 & $3.548 \pm 0.006  $ & $ 273.5 \pm 0.1  $ & $ 1.307 \pm 0.009 $ & 25, 40 \\
                & 2006-11-11 & CAHA 2.2m & LI & Johnson I & $3.553 \pm 0.100  $ & $ 272.5 \pm 1.5  $ & $ 0.750 \pm 0.054 $ & 40 \\
                & 2006-12-26 & VLT/NACO & I & Ks & $3.553 \pm 0.021  $ & $ 273.3 \pm 0.2  $ & $ 1.110 \pm 0.001 $ & 40 \\
                & 2012-09-28 & Gemini North & AO & K & $3.575 \pm 0.010  $ & $ 276.8 \pm 0.2  $ & $ 0.630 \pm 0.041 $ & 37 \\[0.5em]
       GH Tau B & 1991-10-19 & Palomar/Hale 5.1m & SI & K & $0.314 \pm 0.006  $ & $ 119.0 \pm 2.0  $ & $ 0.638 \pm 0.013 $ & 9 \\
                & 1991-11-01$^6$ & CAHA 3.5m & SI & K & $0.350 \pm 0.010  $ & $ 120.0 \pm 1.0  $ & $ 0.102 \pm 0.060 $ & 10 \\
                & 1994-12-19 & Palomar/Hale 5.1m & SI & K & $0.307 \pm 0.002  $ & $ 116.0 \pm 0.4  $ & -- & 14 \\
                & 1996-12-05 & NASA ITF/NSFCAM & SI & K & $0.305 \pm 0.006  $ & $ 114.8 \pm 1.1  $ & $ 0.208 \pm 0.079 $ & 20 \\
                & 1996-12-05  & NASA ITF/NSFCAM & SI & L & $0.312 \pm 0.007  $ & $ 115.3 \pm 1.6  $ & $ 0.372 \pm 0.091 $ & 20 \\
                & 1998-12-05 & HST & I & F814W & $0.311 \pm 0.002  $ & $ 114.6 \pm 0.2  $ & $ 0.084 \pm 0.009 $ & 20 \\
                & 1999-11-17 & Keck & I & N & $0.300 \pm 0.010  $ & $ 111.0 \pm 2.0  $ & $ 0.403 \pm 0.038 $ & 26 \\
                & 2001-01-22 & HST & I & Clear & $0.304 \pm 0.014  $ & $ 113.7 \pm 2.0  $ & -- & 22 \\
                & 2006-11-11 & CAHA 2.2m & LI & Johnson I & $0.286 \pm 0.112  $ & $ 109.0 \pm 21.4  $ & $ 0.378 \pm 0.067 $ & 40 \\
                & 2006-12-26 & VLT/NACO & I & Ks & $0.298 \pm 0.019  $ & $ 109.3 \pm 1.6  $ & $ 0.088 \pm 0.005 $ & 40 \\
                & 2008-12-18 & Keck II & AO & K & $0.299 \pm 0.001  $ & $ 108.5 \pm 0.1  $ & $ 0.079 \pm 0.010 $ & 33 \\
                & 2012-09-28 & Gemini North & AO & K & $0.297 \pm 0.001  $ & $ 106.3 \pm 0.1  $ & $ 0.130 \pm 0.011 $ & 37 \\
                & 2013-01-28 & Keck II & AO & K & $0.295 \pm 0.002  $ & $ 106.2 \pm 0.3  $ & $ 0.118 \pm 0.030 $ & 33 \\[0.5em]
       HN Tau B & 1988-01-15 & CTIO & I & K & $3.100 \pm 0.100  $ & $ 215.0 \pm 2.0  $ & $ 3.430 \pm 0.102 $ & 4 \\
                & 1990-11-26 & San Pedro Martir/0.9m & I & R & $3.100 \pm 0.063  $ & $ 220.0 \pm 1.0  $ & $ 3.090 \pm 0.016 $ & 12 \\
                & 1997-12-06 & NASA ITF/NSFCAM & I & K & $3.109 \pm 0.082  $ & $ 219.1 \pm 1.1  $ & $ 3.721 \pm 0.085 $ & 20 \\
                & 1997-12-06  & NASA ITF/NSFCAM & I & L & $3.150 \pm 0.330  $ & $ 219.2 \pm 2.5  $ & $ 4.532 \pm 0.201 $ & 20 \\
                & 2003-02-18 & VLT/NACO & I & NB\_2.17 & $3.136 \pm 0.006 ^7 $ & $ 219.9 \pm 0.1 ^7 $ & $ 3.603 \pm 0.001 $ & 25, 40 \\
                & 2006-11-12 & CAHA 2.2m & LI & Johnson I & $3.152 \pm 0.084  $ & $ 218.8 \pm 1.5  $ & $ 2.299 \pm 0.056 $ & 40 \\
                & 2012-09-28 & Gemini North & AO & K & $3.210 \pm 0.019  $ & $ 220.0 \pm 0.2  $ & $ 3.090 \pm 0.051 $ & 37 \\[0.5em]
       HV Tau C & 1990-10-08 & IRTF & I & K & $4.000 \pm 0.400  $ & $  45.0 \pm 5.0  $ & $ 3.300 \pm 0.166 $ & 7 \\
                & 1996-08-26 & UKIRT & I & K & $3.980 \pm 0.030  $ & $  43.5 \pm 0.4  $ & $ 4.220 \pm 0.093 $ & 17 \\
                & 1996-08-30 & UKIRT & I & N & $3.990 \pm 0.310  $ & $  44.0 \pm 5.0  $ & $ 0.990 \pm 0.432 $ & 17 \\
                & 1998-02-14 & UKIRT & I & Br$\gamma$ & $4.030 \pm 0.240  $ & $  45.8 \pm 2.0  $ & $ 4.470 \pm 0.224 $ & 17 \\
                & 2000-03-09   & HST & I & F606W & $4.045 \pm 0.071  $ & $  44.1 \pm 1.1  $ & $ 2.979 \pm 0.020 $ & 30, 40 \\
                & 2000-03-09    & HST & I & F814W & $4.044 \pm 0.071  $ & $  43.7 \pm 1.1  $ & $ 2.095 \pm 0.006 $ & 30, 40 \\
                & 2001-12-01 & Keck & I & K & $4.000 \pm 0.200  $ & $  43.0 \pm 1.0  $ & $ 4.663 \pm 0.123 $ & 26 \\
                & 2001-12-01  & Keck & I & L & $3.900 \pm 0.200  $ & $  43.0 \pm 1.0  $ & $ 3.903 \pm 0.054 $ & 26 \\
                & 2002-11-13 & Keck & I & SiC & $4.100 \pm 0.100  $ & $  43.0 \pm 1.0  $ & $ 0.198 \pm 0.010 $ & 26 \\
                & 2002-11-24 & VLT/NACO & I & Ks & $4.001 \pm 0.074  $ & $  44.1 \pm 1.1  $ & $ 4.279 \pm 0.001 $ & 30, 40 \\
                & 2002-11-24  & VLT/NACO & I & H & $3.977 \pm 0.085  $ & $  44.1 \pm 1.3  $ & $ 4.985 \pm 0.001 $ & 30, 40 \\
                & 2002-11-24   & VLT/NACO & I & J & $3.987 \pm 0.109  $ & $  44.0 \pm 1.6  $ & $ 4.944 \pm 0.001 $ & 30, 40 \\
                & 2002-12-13 & Keck II-NIRC2 & I & L & $4.030 \pm 0.010  $ & $  44.9 \pm 0.1  $ & $ 0.689 \pm 0.082 $ & 30 \\
                & 2002-12-13  & Keck II-NIRC2 & I & Ms & $4.070 \pm 0.020  $ & $  44.7 \pm 0.2  $ & -- & 30 \\
                & 2004-12-13  & Keck II-NIRC2 & I & L & $4.020 \pm 0.010  $ & $  44.4 \pm 0.1  $ & $ 0.573 \pm 0.093 $ & 30 \\
                & 2006-11-12 & CAHA 2.2m & LI & Johnson I & $4.129 \pm 0.036  $ & $  42.4 \pm 0.5  $ & $ 4.089 \pm 0.058 $ & 40 \\[0.5em]
     V999 Tau B & 1991-10-27 & CAHA 3.5m & SI & K & $0.300 \pm 0.010  $ & $ 243.0 \pm 20.0  $ & $ 0.555 \pm 0.091 $ & 19 \\
                & 1994-01-27 & CAHA 3.5m & SI & J & $0.260 \pm 0.004  $ & $ 236.4 \pm 0.3  $ & $ 0.848 \pm 0.008 $ & 19 \\
                & 1994-12-15 & CAHA 3.5m & SI & K & $0.252 \pm 0.004  $ & $ 237.5 \pm 1.0  $ & $ 0.689 \pm 0.035 $ & 19 \\
                & 1996-09-28 & CAHA 3.5m & SI & H & $0.239 \pm 0.004  $ & $ 240.1 \pm 0.7  $ & $ 0.965 \pm 0.032 $ & 19 \\
                & 1997-11-17 & CAHA 3.5m & SI & K & $0.234 \pm 0.004  $ & $ 241.6 \pm 0.4  $ & $ 0.831 \pm 0.052 $ & 19 \\
                & 1997-12-08 & NASA ITF/NSFCAM & SI & L & $0.234 \pm 0.005  $ & $ 244.8 \pm 2.9  $ & $ 0.779 \pm 0.101 $ & 20 \\
                & 1998-12-13 & HST & I & Clear & $0.226 \pm 0.014  $ & $ 242.0 \pm 3.1  $ & -- & 22 \\
                & 2006-11-11 & CAHA 2.2m & LI & Johnson I & $0.230 \pm 0.043  $ & $ 278.6 \pm 6.2  $ & $ 0.222 \pm 0.156 $ & 40 \\
                & 2006-12-26 & VLT/NACO & I & Ks & $0.188 \pm 0.004  $ & $ 257.0 \pm 0.6  $ & $ 0.487 \pm 0.003 $ & 40 \\[0.5em]
    V1000 Tau B & 1990-10-04 & Palomar/Hale 5.1m & SI & K & $0.208 \pm 0.002  $ & $  77.6 \pm 0.4  $ & -- & 14 \\
                & 1991-10-27 & CAHA 3.5m & SI & K & $0.230 \pm 0.020  $ & $  85.0 \pm 2.0  $ & $ 0.591 \pm 0.057 $ & 19 \\
                & 1993-11-25 & Palomar/Hale 5.1m & SI & K & $0.230 \pm 0.010  $ & $  83.0 \pm 2.0  $ & -- & 14 \\
                & 1994-10-18 & Palomar/Hale 5.1m & SI & K & $0.224 \pm 0.002  $ & $  82.7 \pm 0.4  $ & -- & 14 \\
                & 1994-12-12 & CAHA 3.5m & SI & K & $0.228 \pm 0.004  $ & $  83.8 \pm 0.3  $ & $ 0.639 \pm 0.012 $ & 19 \\
                & 1996-09-28 & CAHA 3.5m & SI & H & $0.228 \pm 0.004  $ & $  86.3 \pm 0.4  $ & $ 0.630 \pm 0.018 $ & 19 \\
                & 1996-11-29 & CAHA 3.5m & SI & J & $0.236 \pm 0.005  $ & $  87.4 \pm 0.3  $ & $ 0.729 \pm 0.056 $ & 19 \\
                & 1997-03-08 & HST & I & F814W & $0.232 \pm 0.003  $ & $  86.9 \pm 0.4  $ & $ 0.375 \pm 0.010 $ & 20 \\
                & 1997-11-22 & CAHA 3.5m & SI & K & $0.230 \pm 0.004  $ & $  91.0 \pm 0.3  $ & $ 0.731 \pm 0.426 $ & 19 \\
                & 1997-12-08 & NASA ITF/NSFCAM & SI & L & $0.238 \pm 0.005  $ & $  90.7 \pm 1.3  $ & $ 0.602 \pm 0.018 $ & 20 \\
                & 2006-11-11 & CAHA 2.2m & LI & Johnson I & $0.151 \pm 0.049  $ & $  73.4 \pm 14.1  $ & $ 0.739 \pm 0.055 $ & 40 \\
                & 2006-12-26 & VLT/NACO & I & Ks & $0.241 \pm 0.001  $ & $ 100.7 \pm 0.2  $ & $ 0.475 \pm 0.003 $ & 40 \\
                & 2014-12-11 & VLT/SPHERE & XAO & K & $0.235 \pm 0.018  $ & $ 111.3 \pm 2.7  $ & $ 0.592 \pm 0.007 $ & 40 \\[0.5em]
       RW Aur B & 1944-04-01 & McDonald/2.1m & I & Z & $1.220 \pm 0.056  $ & $ 254.3 \pm 0.6  $ & $ 1.500 \pm 0.101 $ & 1 \\
                & 1990-11-09 & Palomar/Hale 5.1m & SI & K & $1.390 \pm 0.030  $ & $ 256.0 \pm 1.0  $ & $ 2.258 \pm 0.136 $ & 9 \\
                & 1991-11-01$^6$ & CAHA 3.5m & SI & K & $1.500 \pm 0.010  $ & $ 258.0 \pm 1.0  $ & $ 1.596 \pm 0.048 $ & 10 \\
                & 1994-11-09 & HST & I & F814W & $1.417 \pm 0.004  $ & $ 255.5 \pm 0.1  $ & $ 2.380 \pm 0.032 $ & 20 \\
                & 1996-12-06 & NASA ITF/NSFCAM & SI & K & $1.397 \pm 0.026  $ & $ 254.6 \pm 1.0  $ & $ 1.566 \pm 0.042 $ & 20 \\
                & 1996-12-06  & NASA ITF/NSFCAM & SI & L & $1.396 \pm 0.026  $ & $ 254.5 \pm 1.0  $ & $ 1.974 \pm 0.043 $ & 20 \\
                & 1999-11-16 & Keck & I & N & $1.420 \pm 0.020  $ & $ 254.6 \pm 0.4  $ & $ 2.836 \pm 0.005 $ & 26 \\
                & 1999-11-16  & Keck & I & IHW18 & $1.380 \pm 0.020  $ & $ 254.7 \pm 0.4  $ & $ 2.593 \pm 0.046 $ & 26 \\
                & 2002-11-17  & VLT/NACO & I & NB\_1.64 & $1.446 \pm 0.005  $ & $ 254.8 \pm 0.2  $ & $ 2.052 \pm 0.003 $ & 25, 40 \\
                & 2002-11-17   & VLT/NACO & I & NB\_2.12 & $1.442 \pm 0.004  $ & $ 255.0 \pm 0.2  $ & $ 1.943 \pm 0.003 $ & 25, 40 \\
                & 2002-11-17    & VLT/NACO & I & NB\_2.17 & $1.444 \pm 0.004  $ & $ 255.1 \pm 0.2  $ & $ 1.990 \pm 0.003 $ & 25, 40 \\
                & 2003-06-01 & IRAM & R & 1.3mm & $1.468 \pm 0.056  $ & $ 255.3 \pm 0.5  $ & $ 1.812 \pm 0.086 $ & 24 \\
                & 2006-11-12 & CAHA 2.2m & LI & RG830 & $1.455 \pm 0.013  $ & $ 253.7 \pm 0.4  $ & $ 1.304 \pm 0.054 $ & 40 \\
                & 2007-06-04 & SAO 6m & SI & V & $1.448 \pm 0.005  $ & $ 255.9 \pm 0.3  $ & $ 1.880 \pm 0.041 $ & 31 \\
                & 2007-06-04  & SAO 6m & SI & I & $1.445 \pm 0.004  $ & $ 255.9 \pm 0.3  $ & $ 1.430 \pm 0.021 $ & 31 \\
                & 2013-01-12 & Chandra & X & -- & $1.480 \pm 0.010  $ & $ 254.3 \pm 0.5  $ & $ 1.175 \pm 0.039 $ & 35 \\[0.5em]
                      
\end{longtable}
\end{longtab}

\nocite{1944PASP...56..123J} \nocite{1989PhDT........11W} \nocite{1990A&A...230L...1H} \nocite{1991A&A...242..428M} \nocite{1991A&A...250..407L} \nocite{1992A&A...261..451B} \nocite{1992ApJ...384..212S} \nocite{1992ESOC...39...47Z} \nocite{1993AJ....106.2005G} \nocite{1993AnA...278..129L} \nocite{1994AnA...283..827T} \nocite{1994ApJ...427..961H} \nocite{1994ApJ...434..707G} \nocite{1995AJ....110..753G} \nocite{1997ApJ...478..766C} \nocite{1997ApJ...481..447K} \nocite{1998A&A...338..122W} \nocite{1999ApJ...515L..35K} \nocite{2001A&A...369..249W} \nocite{2001ApJ...556..265W} \nocite{2003AnA...400..559D} \nocite{2003ApJ...583..334H} \nocite{2003ApJ...588L.113M} \nocite{2006AnA...452..897C} \nocite{2006AnA...459..909C} \nocite{2006ApJ...636..932M} \nocite{2008AJ....135.2496C} \nocite{2008AJ....136.1980K} \nocite{2009ApJ...693L..86C} \nocite{2010ApJ...712..112D} \nocite{2012ARep...56..686B} \nocite{2013ApJ...773...40T} \nocite{2014AJ....147..157S} \nocite{2014ApJ...784...62A} \nocite{2014ApJ...788..101S} \nocite{2014MNRAS.439.4057F} \nocite{2015ApJ...799..155D} \nocite{2015ApJ...808L...3A} \nocite{2016AstL...42...29D} 
\nocite{1993AnA...278...81R}

\section{Data analysis}

In the data analysis we obtained the positions of the stars in the AstraLux, SPHERE and NACO images, calculated their separation and brightness ratios. We combined these results with the other surveys and list them in Table~\ref{table:obs}.

\subsection{Photometry}
\label{sec:psffit}

We performed  PSF photometry on the AstraLux images and aperture photometry on the other images that we analysed to calculate the brightness ratios of the binary pairs. 

In the aperture photometry calculations, we included the following term in the uncertainty of the brightness ratio ($n_{ph}$ is the raw photon count):
\begin{itemize}
  \item $\sqrt{n_{ph}}$ due to the Poisson-process nature of observation;
  \item $\sqrt{\text{aperture area}\times \text{sky background}}$ due to the uncertainty caused by the sky background.
\end{itemize}

In the AstraLux PSF photometry, we used the PSF of lucky images that consists of an Airy disk convolved with a Gaussian and a Moffat function 
(described by \citealt{2010SPIE.7735E.196S}, and successfully used in AstraLux Norte images by \citealt{2014AnA...564A..10W,2015AnA...575A..23W}). This PSF can be expressed as
\[
  \mathrm{PSF}(\vec{r})=W\left(\frac{1}{\vec{r}^2/\sigma_m^2+1}\right)^\beta+\left(1-W\right)\left(\mathrm{PSF}_{\mathrm{th}}(\vec{r}) \cdot e^{\left(\displaystyle -\frac{\vec{r}^2}{2\sigma_g^2}\right)}\right),
\]
where $W$ is a weighting factor between the two components, $\sigma_m$ is the width of the Moffat-profile, $\beta$ is the Moffat power law index and $\sigma_g$ is the width of the Gaussian. We constructed the theoretical AstraLux images as
\[
  \mathrm{AstraLuxImage}(\vec{r}) = \displaystyle \sum_{i=1}^{\mathrm{n_{stars}}} A_i \cdot \mathrm{PSF}_i(\vec{r}) + C_{\mathrm{sky}},
\]
where $A_i$ are the amplitude scaling factors, the $\mathrm{PSF}_i(\vec{r})$ terms are the individual stars,  and $C_{\mathrm{sky}}$ is the sky background. The $\mathrm{PSF}_i(\vec{r})$ terms share the common seeing parameters: $\sigma_m$, $\beta$ and $\sigma_g$, as these parameters do not change in the field of view of the camera with images taken in 20--50~milliseconds. We used the modelling and fitting framework of the Astropy software package \citep{2013A&A...558A..33A} to obtain the PSF fits.

Doing photometry on the AstraLux images differs in a few points compared to the conventional CCD imaging, see e.g. \citet{2010SPIE.7735E.196S}. The main issue affecting the PSF photometry is that the bias level of an electron multiplying CCD (EMCCD) camera can change during acquisition (``bias drift'').
The bias drift can be as high as 0.6\% of the pixel counts. The bias drift can be accounted for by stabilizing the temperature of the camera or by exploiting the overscan region of the camera \citep[see e.g.][]{2012AnA...542A..23H}. Since we did not have any of these options, we calculated the spread of the bias from each night using the sky area of the data cubes. The change in the sky area in the data cubes can be attributed to the bias drift, since no other change is foreseen in the images that were taken 20--50~ms apart. The mean standard deviation of the sky photon count is $0.7\%$, which we took into account as a $1\%$ error term as a safe overestimation of this error.

The final brightness ratios are shown as magnitude differences (with uncertainties) in Table~\ref{table:obs}.

\subsection{Astrometry}

In the case of the SPHERE and NACO observations the analysis of the data is based on the pipeline produced images (including pixel scale and detector position angle), whereas in the case of AstraLux images we calculated the pixel scale and the position angle of the images by using images of the M15 globular cluster \citep{2002AJ....124.3255V} and the Orion Trapezium cluster \citep{2013AJ....146..106O} taken both by the AstraLux system and by the Wide-Field Planetary Camera 2 (WFPC2) on the Hubble Space Telescope (HST). 
We picked 5--9 stars in the clusters which were used to align the AstraLux images and calculated the pixel scale to be $23.71 \pm 0.01$~mas per pixel (this is the pixel scale of the re-sampled and drizzled images which we used in the further analysis; the physical pixel scale is $\approx 47.4$~mas/pixel). The error of this calculation is the root-mean-square deviation of the pixel scale and the position angle of the star pairs.

The position of each companion in the images was obtained by fitting a 2D Gaussian to the stellar profiles utilizing the FITSH software package \citep{2012MNRAS.421.1825P} where the binaries were wide and bright enough to obtain a straightforward fit, and used the PSF fitting (as described in Section~\ref{sec:psffit}) to obtain the coordinates of the close or faint pairs.

The relative position of the stars to each other is also affected by the relative atmospheric diffraction, which we take into account by defining an error term. The possible maximum offset in the relative positions due to this effect is calculated by assuming the maximum relative shift in position of the components diffraction in the used filter (i.e. we take the extreme assumption that the spectral slope of one binary component is very blue while the other is very red, compared to each other). 
We took the wavelengths at the full width at half maximum (FWHM) value of the 
transmission function  of the optical system, where the transmission function is composed from the transmission function of the filter 
(the AstraLux system utilized the RG830\footnote{\url{http://sydor.com/pdfs/Schott_RG830.pdf}}, Johnson I\footnote{\url{http://www.astrodon.com/products/filters/astrodon_photometrics_-_uvbric/}} and SDSS $i'$, $z'$\footnote{\url{http://www.astrodon.com/products/filters/astrodon_photometrics_-_sloan/}} filters) and the quantum efficiency\footnote{The data sheet of the camera used in the AstraLux system is available at: \url{http://www.andor.com/pdfs/specifications/Andor_iXon_897_Specifications.pdf}} of the CCD. Using the airmass at the observed star, we obtained the upper limit of the relative atmospheric diffraction when one or the other of the stars is shifted to the red or blue end of the transmission function.

The astrometric measurements for each epoch of each pair are listed in Table~\ref{table:obs}. The errors listed there are the combined errors due to the uncertainty of the Gaussian fitting of the stellar positions (mean: 2.8 mas), the errors due to the relative atmospheric diffraction (mean: 57.7 mas) and the uncertainty from the pixel scale conversion (mean: 7.8 mas).

\subsection{Orbital fit}
\label{sec:orbitalfit}

We estimated the orbital elements for each companion by fitting models of circular orbits to the astrometric data. A detailed description of the procedure can be found in \citet{2008A&A...482..929K}. In short, it works as follows: to find the period, we employed a grid search in the range 100 to \num{10000} years. For each period, the Thiele-Innes constants were determined using singular value decomposition.  From the Thiele-Innes elements, the semi-major axis $a$, the position angle of the line of nodes $\Omega$, and the inclination $i$ were computed. We have restricted the eccentricity to be zero in all cases as the number of available epochs is moderate and only covers a small portion of the orbit. 
The obtained orbital parameters are listed in Table~\ref{table:rainerfits}.

\begin{table}
\caption{Parameters of the fitted orbits. Eccentricity is zero for all and the $\chi_\nu^2$ is listed in Table~\ref{table:movements}.}
\label{table:rainerfits}
\centering
\begin{tabular}{l|rrrrr}
\hline\hline
Binary pair & $a$ ($''$)& $a$ (AU)  & period (yr) & $\Omega$ ($^\circ$) & $i$ ($^\circ$) \\
\hline
LkCa 3 &  5.8 &  817 & $>\!\num{10000}$ &    161 &     95 \\
DD Tau &  7.4 & 1041 & $>\!\num{10000}$ &     84 &     94 \\
LkCa 7 &  1.1 &  157 &     575 &     27 &     94 \\
FV Tau &  3.6 &  505 & $>\!\num{10000}$ &     49 &     82 \\
FV Tau/c &  0.7 &  103 &     316 &    113 &     86 \\
UX Tau AC &  2.7 &  383 &     692 &      2 &     90 \\
UX Tau AB & 11.0 & 1542 & $>\!\num{10000}$ &    100 &     84 \\
FX Tau &  0.9 &  132 &     417 &    114 &    100 \\
DK Tau &  4.9 &  683 & $>\!\num{10000}$ &    109 &     95 \\
XZ Tau &  0.3 &   42 &     182 &    143 &    116 \\
HK Tau &  8.8 & 1233 & $>\!\num{10000}$ &     88 &    105 \\
V710 Tau &  4.1 &  573 & $>\!\num{10000}$ &     38 &    130 \\
UZ Tau AB &  0.8 &  110 &    1202 &     87 &     62 \\
UZ Tau AC &  5.7 &  796 & $>\!\num{10000}$ &    165 &     53 \\
GH Tau &  0.3 &   48 &     575 &    145 &    138 \\
HN Tau & 11.5 & 1604 & $>\!\num{10000}$ &     51 &     87 \\
HV Tau &  4.0 &  565 &     501 &     44 &     92 \\
V999 Tau & 13.0 & 1826 & $>\!\num{10000}$ &     18 &     89 \\
V1000 Tau &  0.2 &   34 &     138 &     98 &     60 \\
RW Aur &  1.5 &  206 &     955 &     74 &     94 \\

\hline
\multicolumn{6}{l}{\multirow{2}{0.47\textwidth}{
$a$: semi-major axis (using the distance to Taurus star forming region, see Sec.~\ref{sec:distances}),  $\Omega$: position angle of node,  $i$: inclination
}}\\[1.2em]
\hline
\end{tabular}
\end{table}

\subsection{Spectral Energy Distribution}
\label{sec:seds}

To properly characterize the systems in question, we need to know the spectral energy distributions (SED) of the stars. The SED can tell us about the extra or extended emission in the systems, which can be signs of the dust in the disks. 
Since for our purposes the most interesting part of the spectrum is between the optical and millimeter wavelength, we looked for photometric measurement in this region. We employed many photometric surveys: the measurements from \citet{1995ApJS..101..117K,2013ApJ...776...21H} and the IRAS, Spitzer, 2MASS, WISE all-sky surveys, a few ALMA and HST observations along with our lucky imaging data.

The magnitude data were converted to $\lambda F_\lambda$ using the conversion formulas from \citet{1966ARA&A...4..193J,1979PASP...91..589B,1982A&A...107..276G,1992ApJ...398..254B,1974ApJ...191..675G,1988iras....7.....H}, while the data from the surveys were either already given in Jy or had their own conversion formulas to obtain $\lambda F_\lambda$ (e.g. WISE).

We fitted a SED curve on all observations, using the NextGen2 atmospheric models from \citet{1999ApJ...512..377H}\footnote{We used the SEDs available at \url{http://www.am.ub.edu/~carrasco/models/NextGen2/}, with the parameters $M/H=0.0$, $\alpha/Fe=0.0$ and $\log g=4.0$.} and the extinction formulae from \citet{1989ApJ...345..245C}. We adopted $R_V=3.1$ for the total-to-selective extinction ratio, which is the general value for standard interstellar matter. We note, however, that \citet{1985AJ.....90.1490V} also found $R_V=3.1$ applicable to Taurus. 
The spectral types were collected from the literature as listed in Table~\ref{table:spectraltypes}, and we calculated the effective temperatures from the spectral types of each component \citep[using the relations from Table 6 in][]{2013ApJS..208....9P}. We used the calculated effective temperature to select the model of the right temperature from the NextGen2 models.
We fitted the SEDs up to the H-band (except for DD Tau, where we had to use K-band observations due to missing resolved magnitudes at shorter wavelengths) to avoid the influence of the excess emissions from the circumstellar disks present in many systems. The fitted values were the $A_V$ extinction and a scaling value that we obtained using a grid search. 
We have derived the uncertainties of the extinction magnitude using a Monte-Carlo approach by randomly varying the measured flux values within $1\sigma$ and recording the extinction magnitudes of the different runs, repeated a few hundred times.
We employed a ``combined flux'' fitting procedure where we took into account both the individual flux of the components and the combined flux of the whole system at the same time.

\begin{figure*}
\centering
\input{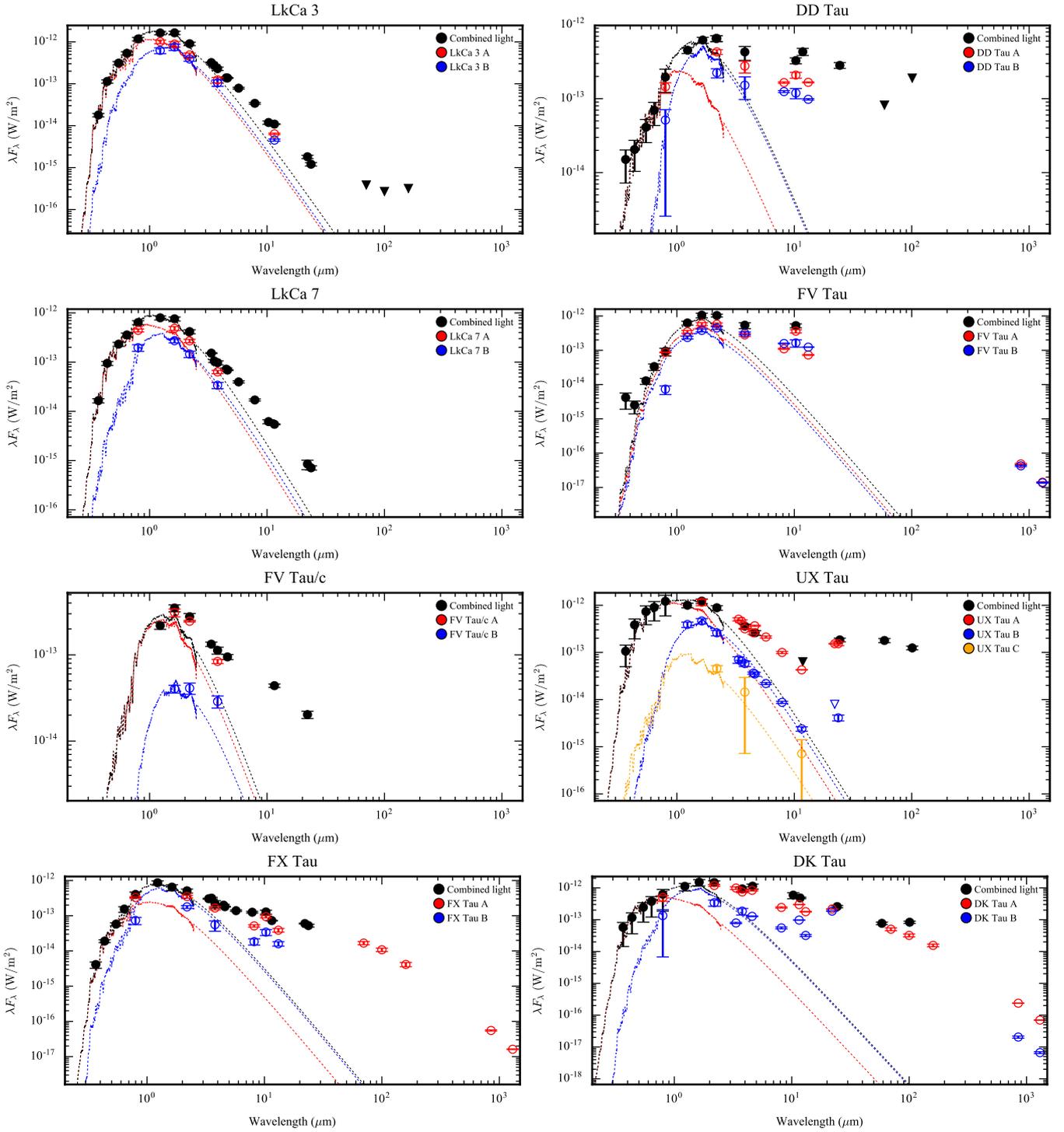}
\caption{SED plots. The triangles show upper limits, the filled markers designate the combined flux from the systems, and the empty markers show the measurements of the individual components. The dotted curves are fitted SED models.}
\label{fig:sed_plots1}
\end{figure*}

The obtained extinction magnitudes are listed in Table~\ref{table:sedparams}, and plots of the fluxes and SED curves are shown in Fig.~\ref{fig:sed_plots1}. The fitted extinctions agree with the literature data within uncertainties for many systems, but there are exceptions. 
In a few systems, only the extinction of the primary is in agreement with the literature (FV Tau, FV Tau/c, HK Tau, V710 Tau, GH Tau and HV Tau), in these cases we anticipate that the discrepancy is attributed to the low number of resolved observations to fit (FV Tau/c, HK Tau, V710 Tau, GH Tau) or that our optical measurement shows a higher extinction for the secondary (in the case of FV Tau) or that we know that the secondary probably has a high extinction magnitude due to an edge-on disk (in the case of HV Tau), where the companion may be seen through their disks.
There are five systems where the obtained extinction magnitudes do not agree with the literature (LkCa 3, XZ Tau, HN Tau, V999 Tau, RW Aur), where we know that two of them show significant variability in time (XZ Tau and RW Aur, see the notes in Appendix~\ref{sec:sample}), one system that is a known quadruple system consisting of two spectroscopic binary systems \citep[LkCa 3,][]{2013ApJ...773...40T}, and there are two systems where the optical fluxes may be the cause of discrepancy (HN Tau and V999 Tau). The latter case, the V999 Tau simply misses reliable optical measurements (although it was observed in the SDSS, it is flagged as a not clean observation), and 
HN Tau features a fairly flat SED in the optical that we cannot achieve a reliable fit using the NextGen2 models.

The NextGen2 models only provide the photospheric SED curve up to 2.5~$\mu$m, therefore to measure the infra-red excess at higher wavelengths, we attached a black body radiation curve to the K-band section of the model curve (with the same effective temperature that we used in the NextGen2 model), making the SED available in the whole wavelength range.

\begin{table}
\setlength{\tabcolsep}{3pt}
\caption{SED fitting parameters. }
\label{table:sedparams}
\centering
\begin{tabular}{lll|lllll}
\hline\hline
Component &   SpT & $T_{eff}^1$ 
 & \multicolumn{1}{l}{$A_V$ (mag)}  & L$^2$ & Mass & \multicolumn{1}{c}{IR ex.$^3$} \\ 
 & & (K) 
 & fit & ($L_\sun$) &($M_\sun$)  & (mag)  \\
\hline
       LkCa 3 A & M2 &   3490  & $\textcolor{black}{0.8}^{+\textcolor{black}{0.8}}_{-\textcolor{black}{0.3}}$ & 0.75 & 0.511 &  1.6  \\[0.20em]
       LkCa 3 B & K7 &   3970  & $\textcolor{black}{3.7}^{+\textcolor{black}{4.3}}_{-\textcolor{black}{2.2}}$ & 0.49 & 1.200 &  1.0  \\[0.55em]
       DD Tau A & M3 &   3360  & $\textcolor{black}{0.7}^{+\textcolor{black}{2.5}}_{-\textcolor{black}{0.7}}$ & 0.15 & 0.274 &  6.8  \\[0.20em]
       DD Tau B & M3 &   3360  & $\textcolor{black}{7.7}^{+\textcolor{black}{0.3}}_{-\textcolor{black}{1.3}}$ & 0.27 & 0.993 &  3.8  \\[0.55em]
       LkCa 7 A & M0 &   3770  & $\textcolor{black}{0.7}^{+\textcolor{black}{0.7}}_{-\textcolor{black}{0.5}}$ & 0.39 & 0.536 & --  \\[0.20em]
       LkCa 7 B & M3.5 &   3260  & $\textcolor{black}{1.7}^{+\textcolor{black}{3.1}}_{-\textcolor{black}{0.8}}$ & 0.23 & 0.259 & --  \\[0.55em]
       FV Tau A & K5 &   4140  & $\textcolor{black}{6.5}^{+\textcolor{black}{1.5}}_{-\textcolor{black}{3.0}}$ & 0.29 & 1.294 &  5.4  \\[0.20em]
       FV Tau B & K6 &   4020  & $\textcolor{black}{8.0}^{+\textcolor{black}{0.0}}_{-\textcolor{black}{0.3}}$ & 0.16 & 1.041 &  4.9  \\[0.55em]
     FV Tau/c A & M2.5 &   3425  & $\textcolor{black}{3.1}^{+\textcolor{black}{0.1}}_{-\textcolor{black}{0.5}}$ & 0.16 & 0.343 & --  \\[0.20em]
     FV Tau/c B & M3.5 &   3260  & $\textcolor{black}{8.0}^{+\textcolor{black}{0.0}}_{-\textcolor{black}{0.9}}$ & 0.03 & 0.234 & --  \\[0.55em]
       UX Tau A & K5 &   4140  & $\textcolor{black}{1.0}^{+\textcolor{black}{2.5}}_{-\textcolor{black}{0.0}}$ & 0.86 & 1.005 &  3.9  \\[0.20em]
       UX Tau B & M2 &   3490  & $\textcolor{black}{7.0}^{+\textcolor{black}{1.0}}_{-\textcolor{black}{7.0}}$ & 0.27 & 1.032 &  0.1  \\[0.20em]
       UX Tau C & M5 &   2880  & $\textcolor{black}{8.0}^{+\textcolor{black}{0.0}}_{-\textcolor{black}{0.0}}$ & 0.01 & 0.087 &  1.5  \\[0.55em]
       FX Tau A & M1 &   3630  & $\textcolor{black}{2.1}^{+\textcolor{black}{0.8}}_{-\textcolor{black}{1.2}}$ & 0.34 & 0.514 &  4.7  \\[0.20em]
       FX Tau B & M4 &   3160  & $\textcolor{black}{4.6}^{+\textcolor{black}{2.8}}_{-\textcolor{black}{2.4}}$ & 0.19 & 0.180 &  3.4  \\[0.55em]
       DK Tau A & K9 &   3880  & $\textcolor{black}{0.0}^{+\textcolor{black}{3.2}}_{-\textcolor{black}{0.0}}$ & 0.36 & 0.552 &  7.2  \\[0.20em]
       DK Tau B & M1 &   3630  & $\textcolor{black}{5.3}^{+\textcolor{black}{2.7}}_{-\textcolor{black}{5.3}}$ & 0.55 & 1.315 &  3.7  \\[0.55em]
       XZ Tau A & M2 &   3490  & $\textcolor{black}{1.1}^{+\textcolor{black}{0.4}}_{-\textcolor{black}{0.2}}$ & 0.15 & 0.334 & --  \\[0.20em]
       XZ Tau B & M3.5 &   3260  & $\textcolor{black}{2.3}^{+\textcolor{black}{0.8}}_{-\textcolor{black}{0.7}}$ & 0.07 & 0.228 & --  \\[0.55em]
       HK Tau A & M1 &   3630  & $\textcolor{black}{2.8}^{+\textcolor{black}{0.4}}_{-\textcolor{black}{0.7}}$ & 0.13 & 0.423 & --  \\[0.20em]
       HK Tau B & M2 &   3490  & $\textcolor{black}{7.6}^{+\textcolor{black}{0.4}}_{-\textcolor{black}{0.7}}$ & 0.03 & 0.334 & --  \\[0.55em]
     V710 Tau A & M0.5 &   3700  & $\textcolor{black}{2.1}^{+\textcolor{black}{0.5}}_{-\textcolor{black}{0.9}}$ & 0.37 & 0.567 & --  \\[0.20em]
     V710 Tau B & M2.5 &   3425  & $\textcolor{black}{8.0}^{+\textcolor{black}{0.0}}_{-\textcolor{black}{4.2}}$ & 0.27 & 1.258 & --  \\[0.55em]
       UZ Tau A & M2 &   3490  & $\textcolor{black}{0.5}^{+\textcolor{black}{2.5}}_{-\textcolor{black}{0.5}}$ & 0.50 & 0.417 &  5.4  \\[0.20em]
       UZ Tau B & M3 &   3360  & $\textcolor{black}{4.0}^{+\textcolor{black}{4.0}}_{-\textcolor{black}{4.0}}$ & 0.35 & 0.482 &  4.5  \\[0.20em]
       UZ Tau C & M1 &   3630  & $\textcolor{black}{4.5}^{+\textcolor{black}{3.5}}_{-\textcolor{black}{4.5}}$ & 0.30 & 0.715 &  3.1  \\[0.55em]
       GH Tau A & M2 &   3490  & $\textcolor{black}{0.6}^{+\textcolor{black}{2.5}}_{-\textcolor{black}{0.6}}$ & 0.29 & 0.360 &  4.5  \\[0.20em]
       GH Tau B & M2 &   3490  & $\textcolor{black}{3.1}^{+\textcolor{black}{4.9}}_{-\textcolor{black}{2.3}}$ & 0.37 & 0.511 &  3.0  \\[0.55em]
       HN Tau A & K5 &   4140  & $\textcolor{black}{3.0}^{+\textcolor{black}{0.2}}_{-\textcolor{black}{1.6}}$ & 0.15 & 0.715 & --  \\[0.20em]
       HN Tau B & M4 &   3160  & $\textcolor{black}{7.6}^{+\textcolor{black}{0.4}}_{-\textcolor{black}{3.2}}$ & 0.02 & 0.190 & --  \\[0.55em]
       HV Tau A & M2 &   3490  & $\textcolor{black}{2.8}^{+\textcolor{black}{0.4}}_{-\textcolor{black}{2.5}}$ & 0.31 & 0.458 &  1.6  \\[0.20em]
       HV Tau C & M0 &   3770  & $\textcolor{black}{8.0}^{+\textcolor{black}{0.0}}_{-\textcolor{black}{0.0}}$ & 0.08 & 0.572 &  2.6  \\[0.55em]
     V999 Tau A & M0.5 &   3700  & $\textcolor{black}{3.2}^{+\textcolor{black}{0.0}}_{-\textcolor{black}{0.4}}$ & 0.27 & 0.567 & --  \\[0.20em]
     V999 Tau B & M2.5 &   3425  & $\textcolor{black}{8.0}^{+\textcolor{black}{0.0}}_{-\textcolor{black}{3.0}}$ & 0.21 & 0.897 & --  \\[0.55em]
    V1000 Tau A & M1 &   3630  & $\textcolor{black}{4.3}^{+\textcolor{black}{1.7}}_{-\textcolor{black}{1.0}}$ & 0.24 & 0.564 & --  \\[0.20em]
    V1000 Tau B & K7 &   3970  & $\textcolor{black}{5.3}^{+\textcolor{black}{2.5}}_{-\textcolor{black}{1.0}}$ & 0.15 & 0.722 & --  \\[0.55em]
       RW Aur A & K2 &   4760  & $\textcolor{black}{3.2}^{+\textcolor{black}{0.0}}_{-\textcolor{black}{0.7}}$ & 0.21 & 1.155 &  6.3  \\[0.20em]
       RW Aur B & K5 &   4140  & $\textcolor{black}{3.3}^{+\textcolor{black}{0.3}}_{-\textcolor{black}{0.8}}$ & 0.81 & 1.294 &  1.6  \\[0.55em]

\hline  
\multicolumn{7}{l}{$^1$ The effective temperature was calculated from the  }\\
\multicolumn{7}{l}{\phantom{$^1$} spectral type and the masses where obtained using}\\
\multicolumn{7}{l}{\phantom{$^1$}  the 2 Myr isochrones from \citet{2015AnA...577A..42B}. }\\
\multicolumn{7}{l}{$^2$ Luminosity,}\\[0.1em]
\multicolumn{7}{l}{$^3$ IR excess at 10~$\mu$m, individual flux, the empty cells (--)}\\[0.1em]
\multicolumn{7}{l}{\phantom{$^1$} indicate stars without resolved 10~$\mu$m observations}\\[0.1em]
\hline  
\end{tabular}
\end{table}

\addtocounter{figure}{-1}
\begin{figure*}
\centering
\input{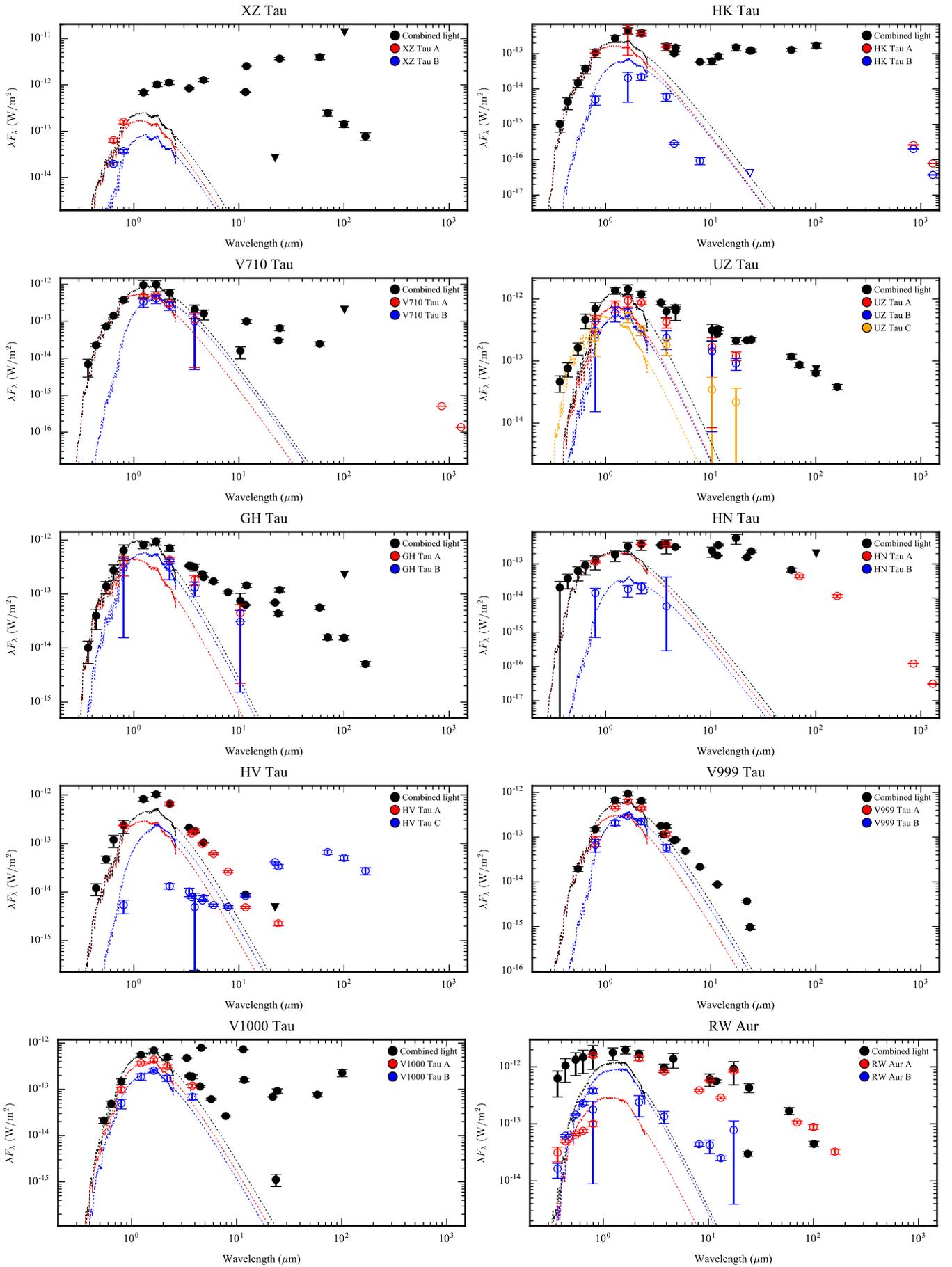}
\caption{SED plots continued.}
\end{figure*}

\section{Results}
\label{sec:results}

\subsection{Astrometry}
\label{sec:astrometry}

\begin{figure*}
\centering
\input{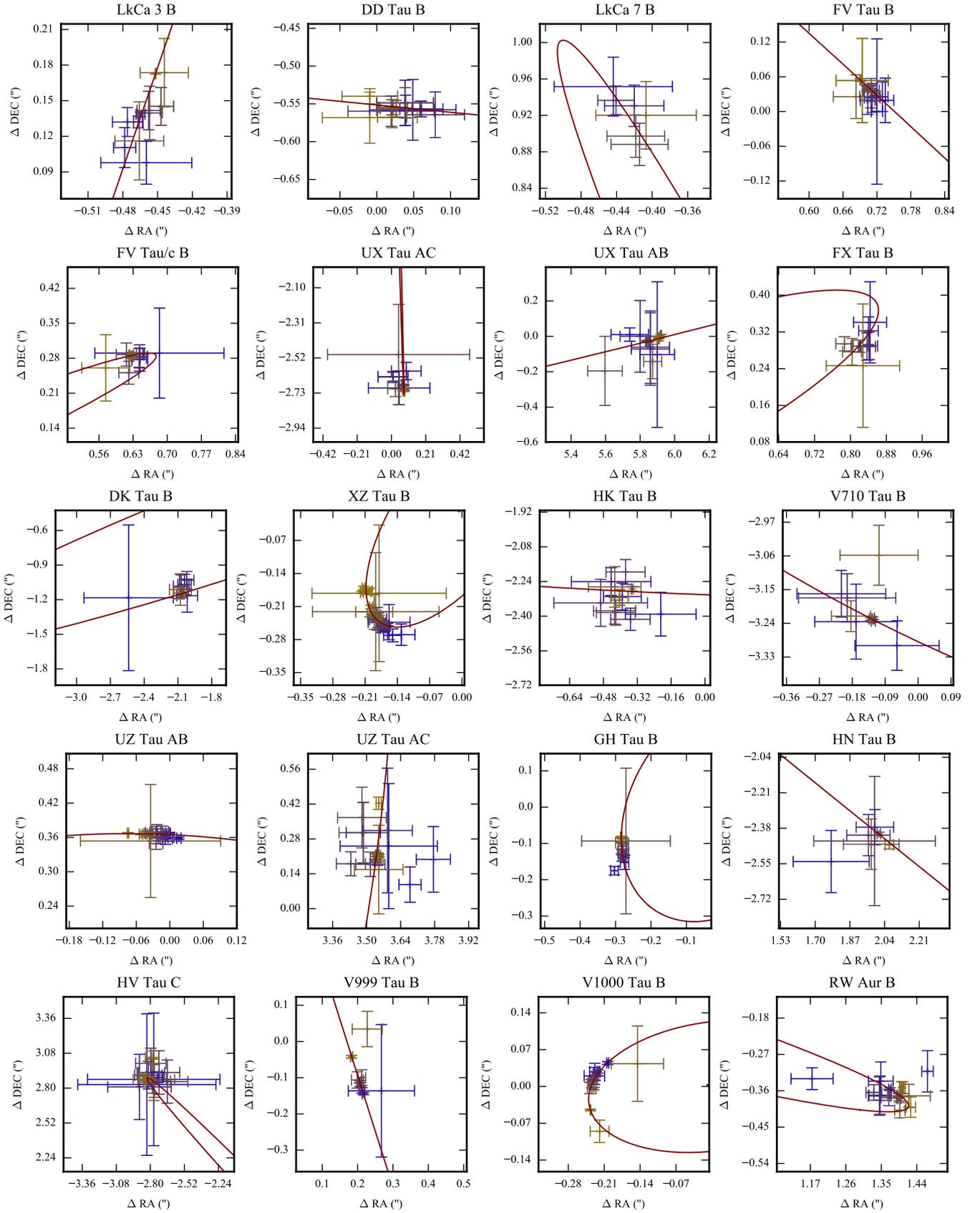}
\caption{Motion of the companions.
The coordinates are relative to the primary star, which is at (0, 0) in the plots (it is marked with a black star symbol when the (0, 0) coordinate is in the plotted area). The blue markers show the oldest epochs, while the brown markers show the latest ones. The maroon curve shows the fitted orbit with $e=0$.}
\label{fig:motions}
\end{figure*}

The individual epochs of the system are shown in Figure~\ref{fig:motions} in RA-DEC plots. The markers of different colours indicate different epochs, the blue ones show the oldest epochs while the brown markers the latest (the epochs are listed in Table~\ref{table:obs}). The dark red curve is a circular, fitted orbit (see details in Section~\ref{sec:orbitalfit}).

\begin{figure*}
\centering
\input{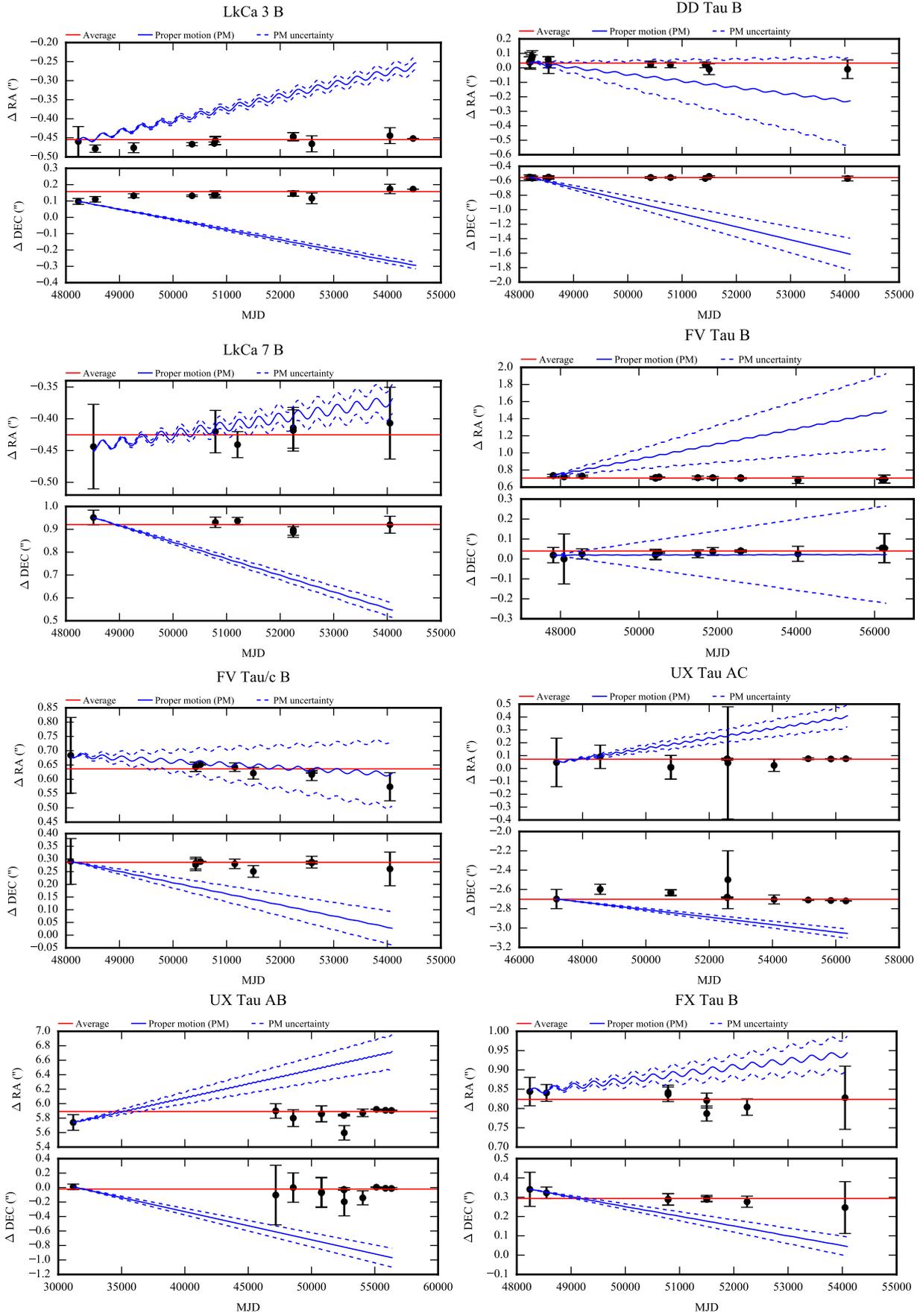}
\caption{Motion of the companions, plotted separately for the RA and DEC dimensions, time dependently. The red line is the average separation, i.e. if the companion is gravitationally bound and has no observable movement, it would line up on the red line. The blue line is the proper motion with the parallax added, the dashed blue line shows the uncertainty of the proper motion. }
\label{fig:timedepepochs1}
\end{figure*}

We also show the individual epochs and the proper motion together in Figure~\ref{fig:timedepepochs1}. Here we separated the RA and DEC coordinates to display the movement of the companion in a (time - spatial dimension) plot. 
The motions of the systems are summarized in Table~\ref{table:movements}. 

\begin{table*}
\setlength{\tabcolsep}{5pt}
\caption{Proper motions, average movements per year, reduced $\chi^2$ of the orbital fits and classification of the binary pairs. 
}
\label{table:movements}
\centering
\begin{tabular}{l|ll|ll|c|r|c}
\hline\hline
Binary pair & \multicolumn{2}{|c|}{Proper motion} & \multicolumn{2}{|c|}{Movement per year} & Relative & Orbital fit, & Orbital${}^1$  \\ \cline{2-5}
 & \multicolumn{1}{c}{RA} & \multicolumn{1}{c|}{DEC} & \multicolumn{1}{c}{RA} & \multicolumn{1}{c|}{DEC}  & motion & \multicolumn{1}{c|}{$\chi_\nu^2$} & motion  \\
 &  \multicolumn{1}{c}{(mas/yr)}&\multicolumn{1}{c|}{(mas/yr)} &\multicolumn{1}{c}{(mas/yr)}&\multicolumn{1}{c|}{(mas/yr)} &  $>3 \sigma$ &  &   \\
\hline
   LkCa 3 & \pms{lkca3}   & \mpy{lkca3b}   & \sigmo{lkca3b}   & \chisqrthree{lkca3b}   & Y \\
   DD Tau & \pms{ddtau}   & \mpy{ddtaub}   & \sigmo{ddtaub}   & \chisqrthree{ddtaub}   & C \\
   LkCa 7 & \pms{lkca7}   & \mpy{lkca7b}   & \sigmo{lkca7b}   & \chisqrthree{lkca7b}   & Y \\
   FV Tau & \pms{fvtau}   & \mpy{fvtaub}   & \sigmo{fvtaub}   & \chisqrthree{fvtaub}   & Y \\
 FV Tau/c & \pms{fvtauc}  & \mpy{fvtaucb}  & \sigmo{fvtaucb}  & \chisqrthree{fvtaucb}  & Y \\
UX Tau AB & \pms{uxtau}   & \mpy{uxtauab}  & \sigmo{uxtauab}  & \chisqrthree{uxtauab}  & N \\
UX Tau AC & \pms{uxtau}   & \mpy{uxtauac}  & \sigmo{uxtauac}  & \chisqrthree{uxtauac}  & Y \\
   FX Tau & \pms{fxtau}   & \mpy{fxtaub}   & \sigmo{fxtaub}   & \chisqrthree{fxtaub}   & C \\
   DK Tau & \pms{dktau}   & \mpy{dktaub}   & \sigmo{dktaub}   & \chisqrthree{dktaub}   & C \\
   XZ Tau & \pms{xztau}   & \mpy{xztaub}   & \sigmo{xztaub}   & \chisqrthree{xztaub}   & Y \\
   HK Tau & \pms{hktau}   & \mpy{hktaub}   & \sigmo{hktaub}   & \chisqrthree{hktaub}   & N \\
 V710 Tau & \pms{v710tau} & \mpy{v710taub} & \sigmo{v710taub} & \chisqrthree{v710taub} & N \\
UZ Tau AB & \pms{uztau}   & \mpy{uztauab}  & \sigmo{uztauab}  & \chisqrthree{uztauab}  & Y \\
UZ Tau AC & \pms{uztau}   & \mpy{uztauac}  & \sigmo{uztauac}  & \chisqrthree{uztauac}  & N \\
   GH Tau & \pms{ghtau}   & \mpy{ghtaub}   & \sigmo{ghtaub}   & \chisqrthree{ghtaub}   & Y \\
   HN Tau & \pms{hntau}   & \mpy{hntaub}   & \sigmo{hntaub}   & \chisqrthree{hntaub}   & C \\
   HV Tau & \pms{hvtau}   & \mpy{hvtauc}   & \sigmo{hvtauc}   & \chisqrthree{hvtauc}   & N \\
 V999 Tau & \pms{v999tau} & \mpy{v999taub} & \sigmo{v999taub} & \chisqrthree{v999taub} & Y \\
V1000 Tau & \pms{v1000tau}& \mpy{v1000taub}& \sigmo{v1000taub}& \chisqrthree{v1000taub} & Y \\
   RW Aur & \pms{rwaur}   & \mpy{rwaurb}   & \sigmo{rwaurb}   & \chisqrthree{rwaurb}   & C \\
\hline
\multicolumn{8}{l}{$^1$ \textit{Y:} orbital motion visible, \textit{C:} orbital motion possible (candidate pair), \textit{N:} orbital motion not seen }\\
\hline
\end{tabular}
\end{table*}

\addtocounter{figure}{-1}
\begin{figure*}
\centering
\input{tables/list_timedep_figs_2.tex}
\caption{Motion of the companions continued.}
\end{figure*}

\addtocounter{figure}{-1}
\begin{figure*}
\centering
\input{tables/list_timedep_figs_3.tex}
\caption{Motion of the companions continued.}
\end{figure*}

If the companion would be a background star, its epochs would line up with the proper motion of the primary star. However, we can see that in all systems, the epochs deviate from the expected proper motion (see the deviation of the black markers from the blue line in Fig.~\ref{fig:timedepepochs1}). 
Since none of our binaries show motion comparable with the proper motion, we can conclude that each pair is either a common proper motion pair or a gravitationally bound pair. 

\subsection{Orbital fit}

\begin{figure*}
\centering
\input{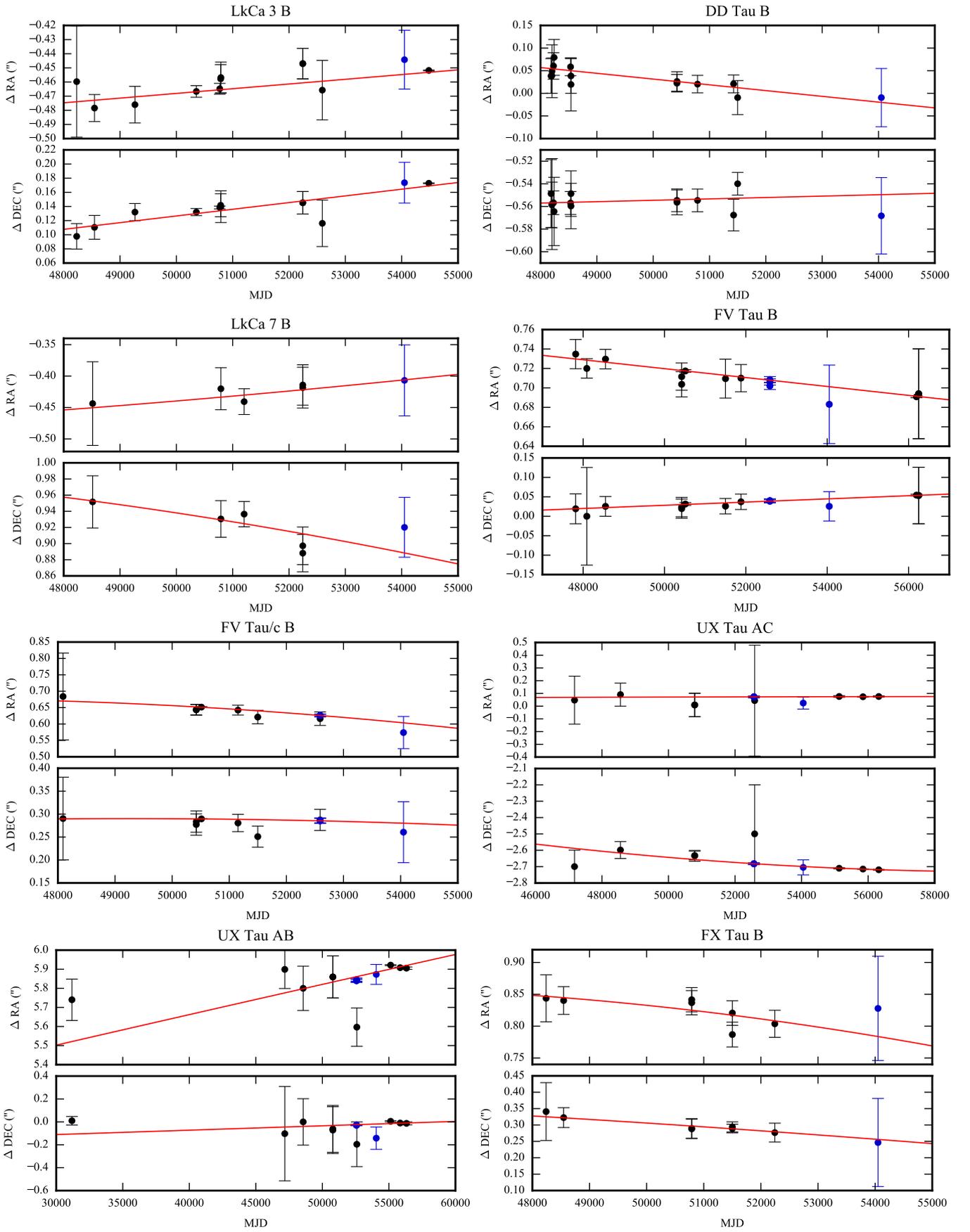}
\caption{Motion of the companions, plotted separately for the RA and DEC dimensions, time dependently. The red line is the orbit from the orbital fit. The black markers are literature data points, the blue ones are data points from our observations. }
\label{fig:temorb1}
\end{figure*}

We looked at the relative motion of the companion around the primary by obtaining orbital fits to decide whether the pair is a common proper motion pair or is gravitationally bound (see Section~\ref{sec:orbitalfit} for the description of the fitting procedure).
We calculated the average relative motion of the companions and compiled Table~\ref{table:movements} to convey the following classifications: \textit{relative motion over $3 \sigma$,} which shows whether there is at least a 3 $\sigma$ difference between any two astronomical epochs, i.e. there is detectable relative motion; and \textit{orbital motion,} which is a classification of the binary pair based on the $\chi_\nu^2$ value of the orbital fit and the visual inspection in Fig.~\ref{fig:temorb1}. 

We based our visual inspection on the fact that if the companion is orbiting the primary then its coordinates must show a rising or decreasing trend in either RA or DEC (or both), in Fig.~\ref{fig:temorb1}. If we do not see such a trend then the companion is likely to move with the same proper motion as the primary star. The classification letters in Table~\ref{table:movements} indicate the pairs orbiting each other with 'Y' (which are also gravitationally bound); the ones that likely are orbiting each other, but the observational uncertainties are too high to draw certain conclusions with 'C'; and the ones which do not show orbital motion with 'N'. The last group likely contains common proper motion pairs, because we do not see any pairs where the companion would be a background star (as mentioned in Section~\ref{sec:astrometry} and shown in Fig.~\ref{fig:timedepepochs1}).
Since stars in the same star forming region may have a similar evolutionary history, it is possible to find several binary pairs which have similar proper motions without being gravitationally bound. As for the statistics, there are ten pairs with detected orbital motion, five with possible orbital motion and five which are most likely common proper motion pairs.

\addtocounter{figure}{-1}
\begin{figure*}
\centering
\input{tables/list_temorb_figs_2.tex}
\caption{Motion of the companions with fitted orbit, continued.}
\end{figure*}

\addtocounter{figure}{-1}
\begin{figure*}
\centering
\input{tables/list_temorb_figs_3.tex}
\caption{Motion of the companions with fitted orbit, continued.}
\end{figure*}

\subsection{Stellar parameters and disk statistics}

We calculated several properties from the SED fits. In Table~\ref{table:sedparams} we list the parameters of the SEDs (spectral type, effective temperature, extinction) and the derived parameters (luminosity, mass and IR excess). 

We calculated the stellar masses by using the 2 Myr isochrones from \citet{2015AnA...577A..42B}, where we utilized the effective temperature and the luminosity as input parameters from Section~\ref{sec:seds}. The derived masses of the components vary mainly between 0.08 and 1.32~$M_\sun$ which agrees with the general assumption that T Tauri stars are low mass stars. 

Infrared excesses were calculated at 10~$\mu$m, using either N-band, W3 (WISE) or IRAS 11.8~$\mu$m  measurements. The difference is calculated between the fitted SED curve (which depends on the effective temperature, and therefore, on the spectral type) and the individual photometric data points, i.e. we calculate the excess over the photospheric level.  
The IR excesses of the individual stars vary between 0.1 and 7.2 magnitudes.

There are two resolved triple systems in our sample, UX Tau and UZ Tau. Unlike binary systems, triple systems can be unstable in some configurations. We looked at two stability indicators, the separations and the masses. 

\citet{2015AJ....149..145R} have examined the stability of triple systems based on the masses of the companions. Their 'triple diagnostic  diagram' is based on the $M_a/M_b$ and $M_c/(M_a+M_b+M_c)$ ratios of the systems. Based on our derived stellar masses the ratios in the UX Tau system are: $M_b/M_a=0.97$ and $M_c/(M_a+M_b+M_c)=0.04$. In the UZ Tau system, they are: $M_b/M_a=0.67$ and $M_c/(M_a+M_b+M_c)=0.26$. These ratios place both the UX Tau and the UZ Tau system in the disrupted regime of the triple diagnostic diagram.

It is well known that triple systems, in which the third body is closer than $\sim10$ times the separation of the other two bodies, are inherently unstable \citep{2014prpl.conf..267R}. 
In UX Tau, the semi-major axes are 383~AU and 1541~AU, while in the other triple system, UZ Tau, the semi-major axes are 110~AU and 797~AU. These values would indicate an unstable configuration in both cases. Moreover, we found that UX Tau AB and UZ Tau AC are likely gravitationally bound, but UX Tau AC and UZ Tau AB do not show orbital motion. Therefore it is possible that these systems are not physical triple systems, but only binary pairs coupled with common proper motion stars. However, since the fraction of the orbits covered by the observations so far is small, the determined semi-major axes may be significantly affected by the inclination of the orbits to our line of sight, thus we cannot confirm the classification of either UX Tau or UZ Tau as a stable or unstable triple system.

\begin{figure*}
\centering
\includegraphics[width=0.30\textwidth]{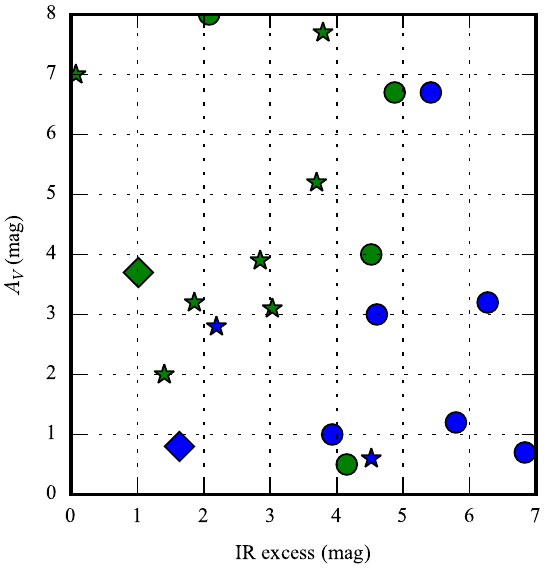}
\includegraphics[width=0.30\textwidth]{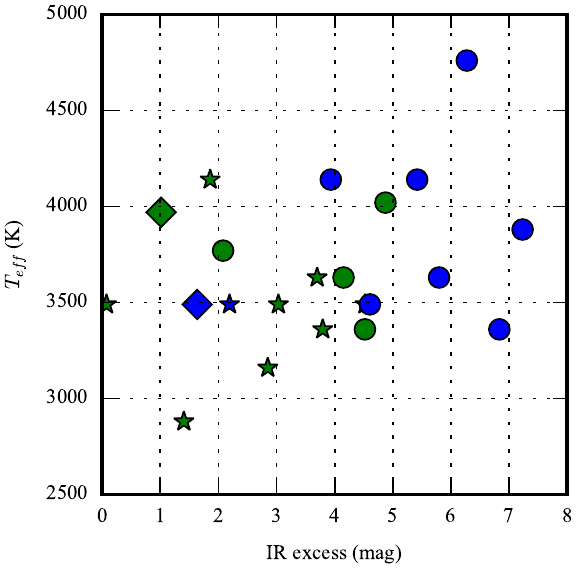}
\includegraphics[width=0.30\textwidth]{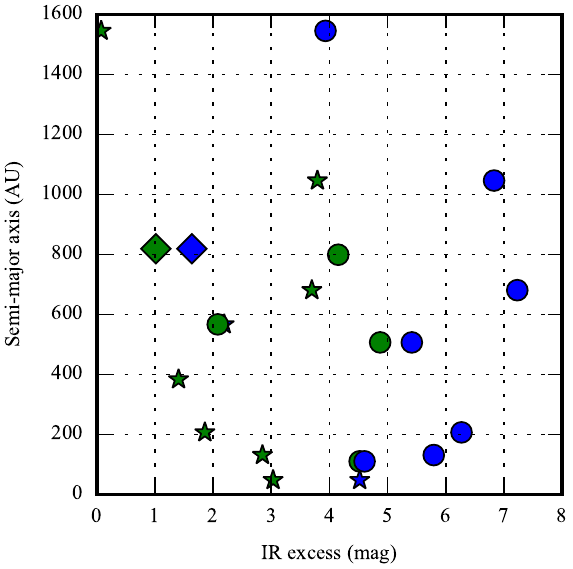}
\caption{Scatter correlation plots. 
The filled circles show stars with disks observed at long wavelengths, the star markers stand for the stars without a disk detected at long wavelengths (however, we stress that a disk could be still present at detection levels lower than what was used by \citet{2012ApJ...751..115H} or \citet{2006AnA...452..897C})
and the diamonds are the stars where we have no high spatial resolution long wavelength observation of the systems.
The blue markers are primary, the green markers are secondary/tertiary stars.}
\label{fig:scatterplots}
\end{figure*}

\section{Discussion}
\label{sec:discussion}

We plotted the derived stellar and binary parameters as a function of the IR excesses in Figure~\ref{fig:scatterplots}. The star markers show stars around which the long wavelength observations  did not resolve a disk, the filled circles show the stars where these observations did resolve a disk and the diamonds are the stars which do not have long wavelength measurements. The long wavelength observations probe the emission from the outer disk \citep[and references therein]{2014prpl.conf.....B}, while the IR excess at 10~$\mu$m shows the emission from the inner disk. 

In the Figure~\ref{fig:scatterplots} (left panel, IR excess - $A_V$) there are stars in all quadrants in this projection of the parameter space, and we do not see correlation. 
The middle and right panels in Figure~\ref{fig:scatterplots} (IR excess - T$_{eff}$ and IR excess - Semi-major axis) also shows no correlations, which also applies to the mass and luminosity as a function of IR excess.

In the IR excess plots we can see that all but one star with disks detected at long wavelengths have 10~$\mu$m IR excess over 3.9 magnitudes. From the opposite direction, we can also see that all but one stars with IR excesses over 3.9 magnitudes have a disk detected at longer wavelengths. In the first case the exception is HV Tau C, which is a star with an edge-on disk, and its T Tauri classification falls between Class I and II due to its flat SED around 10~$\mu$m. Its IR excess is only moderate although it definitely harbours a disk, but the edge-on disk may decrease the detectable IR excess.
In the second case the exception is GH Tau A, which has an IR excess of 4.5 magnitude, but we note that the uncertainties of its 10~$\mu$m measurements are high, therefore it may be labelled as an outlier. We also note that the intrinsic variability of T Tauri stars can be significant, and the outliers might be attributed to the fact that many of the photometric measurements were taken apart in time.

In a binary pair of young stars it is possible that one star has an influence on the disk of the other star, and such influence may be visible in the emission originating from the disk. Therefore we examined the distribution of IR excess in single T Tauri stars to check if we can see any difference between the distribution of the IR excess of single stars and stars in multiple systems.
We selected 40 single stars which have full UVBRI photometry, have no sign of multiplicity and present a similar spectral type distribution as the binary sample. We have selected the single stars from the \citet{1995ApJS..101..117K}, \citet{2004AJ....128..805S} and \citet{2017AJ....153...46L} papers, among six other stars found in SIMBAD by querying  the Taurus-Auriga region for T Tauri objects of specific spectral types (listed in Table~\ref{table:singlessedparams}). To ensure that the selected stars are single, we checked the stars in SIMBAD, Vizier and the Washington Double Star Catalog (WDS), looking for any sign of multiplicity, such as dual SED curves, dual spectral types, notes on duality or components present in the WDS with a separation smaller than 5.8\arcsec, the upper limit of separations in our sample. The resulting sample have been analysed similarly to our multiple system: we added 10~$\mu$m photometry data from the WISE survey, fitted the SED curves and derived the IR excess.

The 10~\um IR excess is measured in the Rayleigh-Jeans region of the SED, which is not as dependent on the effective temperature as e.g. the K-band magnitudes, but it is still affected by it. 

\begin{figure}
\centering
\includegraphics[width=0.45\textwidth]{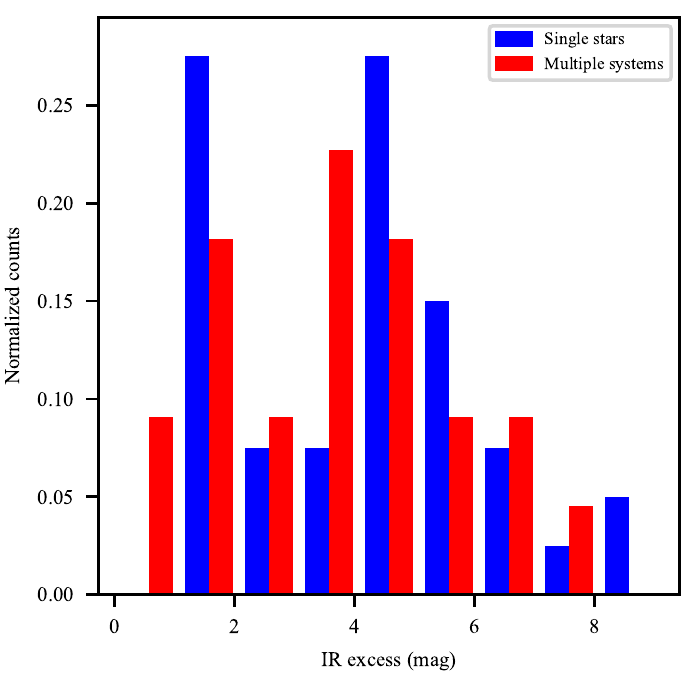}
\caption{
10~$\mu$m IR excesses by multiplicity. \emph{Blue:} Single stars, \emph{Red:} Multiple systems (individual flux measurements).
}
\label{fig:ir_excesses_histograms}
\end{figure}

The results of the comparison of the 10~\um excesses is plotted in Figure~\ref{fig:ir_excesses_histograms}, the IR excesses of the single stars are between 1.7 and 9.1 magnitudes. The main difference between the two distributions is that the multiple sample has slightly more stars with high IR excesses while the single sample has more stars with low IR excesses. We performed a Kolmogorov-Smirnov test to compare the two distributions of the IR excesses. The resulting  $D_{KS} = 0.27$ difference and $p_{KS} = 0.20$ probability indicates that the null hypothesis (in which the underlying distributions would be identical) cannot be rejected at a value of $p < 0.05$ (and a confidence of ${} > 0.95$). Therefore, although one could expect that the presence of a companion could affect the emission coming from the inner disk, the apparent difference between the IR excess of the single and multiple objects may be due to statistical fluctuations.

\section{Conclusions}

We have carried out a survey of 18 multiple T Tauri systems with the goal to detect orbital motion, determine orbital parameters and also look at correlations between the binary configuration, the disk state and stellar parameters. Our sample covers binary separations from 0.22\arcsec to 5.8\arcsec, and spectral types K1 to M5 (corresponding to masses between 0.08 and 1.32~$M_\sun$).

We found that ten pairs out of 20 are orbiting each other, five pairs may show orbital motion, and five are likely common proper motion pairs. We found no obvious correlation between the stellar parameters and binary configuration. The 10~$\mu$m IR excess of the multiple systems varies between 0.1 and 7.2 magnitudes, while it is between 1.7 and 9.1 in the sample of single stars. The distribution of the IR excesses in the two samples provide no statistical evidence for being different distributions, therefore the presence of the companion does not affect the emission coming from the inner disk.

We note that we have not detected any signs of circumbinary disks, which usually are also present in young multiple stellar system. However, they are more abundant around very close binaries with a separation of a few AU \citep{2014prpl.conf..267R}, which is not the case for our sample. Also, the circumbinary disks have to have larger inner holes due to the central stars (in our sample the holes would be larger than 30 AU), therefore their infrared excess can be easily below the sensitivity of our observations.

The obtained orbital periods vary from 138 year to over \num{10000} years. This suggests that even for the shortest orbital periods, the observation from the last $\approx 20$ years may only cover $\approx 15$\% of the orbit, therefore obtaining precise orbital fits is yet not viable. We may be able to obtain a meaningful astrometric orbit by re-observing seven of the systems included in our study in 15 years.
 
\begin{acknowledgements} 
The authors would like to thank the SPHERE Science Verification team for their support in obtaining the observations.
This work was supported by the Hungarian Scholarship Board Office.
This work was supported by the Hungarian Academy of Sciences via the grant LP2012-31.
This work was supported by the Momentum grant of the MTA CSFK Lend\"ulet Disk Research Group, and the Hungarian Research Fund OTKA grant K101393.
This research has made use of the Washington Double Star Catalog maintained at the U.S. Naval Observatory.
Based on observations made at the La Silla Paranal Observatory under programme ID 60.A-9364(A), 70.C-0565(A), 70.C-0701(A), 72.C-0022(A), and 78.C-0386(A).
\end{acknowledgements}

\bibliographystyle{aa}
\bibliography{ttbs}

\appendix

\section{Description of the individual targets}
\label{sec:sample}

\paragraph{\object{LkCa 3}} This system was long known as a visual binary \citep[since it has been resolved by][]{2001A&A...369..249W}, but recent observations from \citet{2013ApJ...773...40T} have shown that it is actually a hierarchical quadruple system of M-type stars. With the assumption of coevality, they found an age of 1.4 Myr and a distance of 133 pc for the system, which is consistent with previous estimates for the region of LkCa 3 and suggests that the system is on the near side of the Taurus complex. 

\paragraph{\object{DD Tau}} was resolved as a binary by \citet{1992AA...261..451B}, who have obtained high-resolution images of the T Tauri star DD Tau at optical and near-IR wavelengths, between 0.5 and 3.9 $\mu m$. The system was resolved into two intensity peaks, with a separation of $0.56 \pm 0.01$\arcsec and position angle (PA) of $184 \pm 2^\circ$. The photometry of the two components suggested that DD Tau is a binary system consisting of two active T Tauri stars of similar luminosity. They also found evidence for an extended nebulosity, which may be a reflection nebula illuminated by the two stars.
\citet{1992ApJ...397L.107G} presented high resolution and narrow-band images of DD Tau, obtained with the High-Resolution Camera of CFHT in which they also resolved the forbidden line emission region. They found that the bulk of the continuum emission is concentrated in two knots that are separated by 0.55\arcsec and oriented in a PA of $6.4^\circ$. 
The forbidden-line morphology was found to be distinctly different from that of the continuum. The forbidden N\,{\sc ii} emission appeared to be a 'jetlike' extension connecting the two continuum knots, while the forbidden S\,{\sc ii} emission seems to be confined to the northern knot. The southern knot (DD Tau B) is extended in the of PA $\approx 130^\circ$. The orientation of the disk as defined by IR polarization measurements is PA $\approx 125^\circ$.

\paragraph{\object{LkCa 7}} The high resolution survey by \citet{1993AnA...278..129L} showed that it is a binary system. The primary is known the be a weak-line T Tauri star, as observed by e.g. \citet{1998AstL...24...48G}. It was part of a long term variability survey \citep{2008A&A...479..827G}, where they measured the stellar rotation period.

\paragraph{FV Tau and FV Tau/c}  These two binary systems are separated by $12.29\arcsec$ \citep{2009ApJ...703.1511K}. They are usually included in the T Tauri binary surveys, and referred as a wide binary pair, probably forming a quadruple system \citep[see e.g.][]{1993AJ....106.2005G,2003ApJ...583..334H}. However, there is no evidence that they are gravitationally bound, thus we handle them as two separate systems. Since their separation is too large for most of the high spatial resolution instruments that we used, combining the separations from two different pointings would lead to higher astrometric uncertainties. Therefore our data does not allow us to test whether FV Tau and FV Tau/c indeed form a common proper motion quadruple system.

\paragraph{\object{FV Tau}} This system was first resolved by \citet{1990ApJ...357..224C} using lunar occultation measurements, but \citet{1991A&A...250..407L} measured the binary parameters first. FV Tau (and FV Tau/c) was included in a recent survey conducted by \citet{2014ApJ...784...62A}, who used ALMA at two wavelengths (850 $\mu$m and 1.3 mm). They derived the stellar and disk masses of the primary, $1.2^{+0.21}_{-0.42}\ M_\sun$ and 
$\left(6.3 \pm 1.4 \right) \cdot 10^{-4}\ M_\sun$, respectively.

\paragraph{FV Tau/c (also as \object{HBC 387})} \citet{1992ApJ...384..212S} resolved this binary. As its name suggests, it is very close to FV Tau (12.3\arcsec) and was also included in the survey by \citet{2014ApJ...784...62A} (using ALMA at 850 $\mu$m and 1.3 mm). However, the measured separation does not agree with the previous observations (the discrepancy is $\approx 3 \sigma$ in the separation, suggesting another body that emits the radio), therefore we did not include this epoch in our analysis.

\paragraph{\object{UX Tau}} The two brightest components (A and B) in UX Tau were already resolved in 1944 by \citet{1944PASP...56..123J}. The third component (C) was seen by Herbig in 1975 \citep{1979AJ.....84.1872J}, but \citet{1992RMxAA..24..109D} were the first to measure the actual separation. The forth component (D) was discovered by \citet{1999A&A...341..547D} at CFHT, using adaptive optics.

\citet{2007ApJ...670L.135E} have analysed the Spitzer IRS spectra for UX Tau A and their SED fittings suggested the existence of a disk gap of $\approx 56$~AU. A few years later \citet{2010ApJ...717..441E} measured the gap to be 71~AU wide, using near-infrared spectral measurements. \citet{2012PASJ...64..124T} also examined the disk around UX Tau A, using near-infra-red observations from the Subaru telescope, who found 
a strongly polarized circumstellar disk surrounding UX Tau A and extending to 120 AU, at a spatial resolution of 0.1\arcsec (14 AU). The disk is inclined by $46^\circ \pm 2^\circ$, with the west side being the nearest. They have not detected the gap that was suggested by SED models at the limit of their inner working angle (23 AU) at the near-infrared wavelength.

UX Tau C was observed by \citet{2003ApJ...582.1109W} using the Keck I telescope to obtain high resolution spectra. They re-determined the spectral type to be M5 and calculated the mass to be $0.166 \pm 0.047 M_\sun$. \citet{2011ApJ...732...42A} found no evidence for remnant disk material, nor detected 880$\mu$m emission. This is in agreement with \citet{2006ApJ...636..932M} who determined UX Tau A to be a classical T Tauri star, while UX Tau B and C are weak line T Tauris.

\paragraph{\object{FX Tau}} This system was also resolved by \citet{1993AnA...278..129L}. \citet{2014ApJ...784...62A} have observed this system with ALMA, but did not detect the companion.

\paragraph{\object{DK Tau}} The first observer of this binary system was \citet{1989PhDT........11W}, using speckle imaging. \citet{1992ApJ...384..212S} also resolved the binary using occultation and optical measurements, but the position they report deviates with more than 5~$\sigma$ from the other astrometric observations. However, their numbers strongly suggest a coordinate conversion error (a sign error in the right ascension H:M:S $\to$ degree conversion), thus we revisited their data and recalculated the astrometric position. Recently, the system was resolved by \citet{2014ApJ...784...62A} using ALMA, but since the separation does not agree with the positions therein, we recalculated the separation based the RA and DEC coordinates reported there.

\paragraph{\object{XZ Tau}} \citet{1990A&A...230L...1H} resolved XZ Tau as a binary system using near-infrared speckle observations.
\citet{1997ApJ...478..766C} have resolved the system and performed astrometric and photometric measurements. However, since their focus was HL Tau, they only provided coarse photometry for XZ Tau which includes outflows around the system. 
\citet{2008AJ....136.1980K} have monitored the system and its bipolar outflow over ten years using HST. They  found traces of shocked emission as far as 20\arcsec south of the binary.
VLA observations carried out by \citet{2009ApJ...693L..86C} resolved the southern component into a double system with a separation of 90~mas, making the XZ Tau system a triple. The third component is likely deeply embedded and not visible at optical and IR wavelengths. \citet{2014MNRAS.439.4057F} have revisited the possibility of a third component by observing the system again with VLA, but they found no trace of that third component.
\citet{2016AstL...42...29D} have made recent observation of the system using a 6-m class telescope with speckle imaging and also derived preliminary orbital parameters.

\paragraph{\object{HK Tau}} was first resolved by \citet{1991A&A...242..428M} using classical infra-red imaging. It was also included in the ALMA survey by \citet{2014ApJ...784...62A}.
It is known to be the first young binary system where an edge-on disk was found using HST  \citep{1998ApJ...502L..65S,2011ApJ...727...90M}. The secondary star is heavily extincted due to its disk, and, in optical and IR, only the scattered light is detectable from that companion.
Later observations revealed that the primary also harbours a disk, making this system a young pre-main-sequence binary with two circumstellar disks \citep{2012ApJ...751..115H}.

\paragraph{\object{V710 Tau}} This system was first resolved by \citet{1979ApJS...41..743C} using infra-red spectroscopy, and the binary parameters were determined later by \citet{1993AnA...278..129L} and \citet{1994ApJ...427..961H}. It is a classic and weak-line T Tauri star pair, the primary is slightly more massive than the secondary \citep{2001ApJ...556..265W,2003ApJ...584..875J}.  \citet{2008ApJ...683..893S} spatially resolved the system using Chandra and they found that both components emit X-rays.
V710 Tau was observed by \citet{2014ApJ...784...62A} as well using ALMA, but they could only detect the primary star.

\paragraph{\object{UZ Tau}} The two brightest star of this multiple system were already noted by \citet{1944PASP...56..123J} and later the system was resolved as a triple system by \citet{1992ApJ...384..212S}. The UZ Tau W binary was extensively examined by \citet{1996AJ....111.2431J}, and in the same year, \citet{1996BAAS...28S.920M} have obtained spectroscopic measurements of UZ Tau E which showed it to be a spectroscopic binary, making the system quadruple. The UZ Tau E binary was closely examined by \citet{2007AJ....134..241J} who found that the brightness varies with a period of $19.16 \pm 0.04$ days which is consistent with the previous spectroscopic binary period of 19.13 days. Their best orbital fit resulted in a separation of $0.124 \pm 0.003$ AU which converts to less than 1~mas spatial separation, making it as of yet an unresolvable binary.

\paragraph{\object{GH Tau}} The GH Tau binary was resolved first by \citet{1993AnA...278..129L}. This binary was included in the surveys later by \citet{2006ApJ...636..932M} and \citet{2003ApJ...583..334H}.

\paragraph{\object{HN Tau}} HN Tau was resolved as a binary system by \citet{1991A&A...242..428M}. It was also observed in the ALMA survey by \citet{2014ApJ...784...62A}, but they did not detect the secondary.

\paragraph{\object{HV Tau}} This is a triple system, resolved first by \citet{1992ApJ...384..212S}. HV Tau A-B is a close binary (74~mas, measured in 1996) with similar brightnesses, therefore in many surveys only the HV Tau AB--C pair is resolved. The HV Tau A-B pair was recently re-observed by \citet{2011ApJ...731....8K} using the Keck telescope, and they reported $36.0 \pm 0.2$~mas for the separation and $326.6 \pm 0.3$ degree for the position angle.
The system was closely examined by \citet{2010ApJ...712..112D} using NACO at the VLT.
They found that HV Tau AB--C is a common proper motion pair and that the orbital motion within the close HV Tau AB system is slow, suggesting a highly eccentric orbit or a large
de-projected physical separation. Previous spectroscopic and photometric measurements also showed that the AB subsystem does not experience accretion nor does it show infra-red excess \citep{2001ApJ...556..265W}. HV Tau C star has an almost edge-on disk, which has a mass in the range of $M_{disk} \sim 10^{-3}M_\sun$, and $R_{out}=50$~AU size \citep{2010ApJ...712..112D}. 
  
The T Tauri classification of HV Tau C is ambiguous, it is usually classified as ``I?'' because the SED is rising in the NIR regime, but it levels out after $\approx 20\ \mu$m. Due to its almost edge-on and optically thick disk which totally block out the star as a point source, \citet{2003ApJ...589..410S} suggested that an extinction magnitude of $A_V>50$ shall be adopted for this system. 
Due to the edge-on disk, we are also unable to precisely locate the position of the star, which can be seen in the positional plot in Figure~\ref{fig:motions}, where the astrometric epochs overlap each other at the assumed position of the star. This makes the possible orbital fit difficult, and we could either need to measure the exact position of the star more accurately or wait for the companion to orbit around the primary for a few more decades before deriving precise orbital parameters for this system.

\paragraph{\object{V955 Tau}, \object{V999 Tau}, \object{V1000 Tau}} These stars are also known as LkH$\alpha$ 332, LkH$\alpha$ 332 G2, LkH$\alpha$ 332 G1, respectively. They are three close binary pairs which are $\approx 11$\arcsec and $\approx 26$\arcsec away from each other, probably forming a wide triple system \citep{2009ApJ...703.1511K}. All three stars were resolved to be binaries by \citet{2001A&A...369..249W} and V1000 Tau was a bit earlier resolved by \citet{1995AJ....110..753G}. \citet{2003ApJ...583..334H} have determined V955 Tau to be a classical T Tauri pair, while V999 Tau is a weak-line T Tauri pair. 

\paragraph{\object{RW Aur}} This star was resolved by \citet{1944PASP...56..123J} as a binary, and it is also shown  as a binary pair in the third Herbig-Bell catalogue \citep{1988cels.book.....H}. 
\citet{2006AnA...452..897C} have resolved an optically thick  disk around the primary which have a radius of 40--57~AU. The primary also features one of the highest known accretion rates for a T Tauri star \citep{1995ApJ...452..736H}.
The primary star features an optically visible, asymmetric bipolar outflow, and was recently resolved by \citet{2014ApJ...788..101S} using Chandra. They found that both components are visible in X-rays, and the luminosity of the less-massive secondary is at least twice that of the primary and is variable. This binary also features a peculiar dynamical configuration: the disk around RW Aur A is counter-rotating with respect to the binary orbital motion \citep{2012ARep...56..686B}.
It is a significantly variable system, \citet{2013AJ....146..112R} have observed a dimming that had a depth of $\approx 2$~mag and a duration of $\approx 180$~days between 2010 September and 2011 March. They speculated that the dimming may be attributed to a one-time occultation by the tidally disrupted disk around the primary.

\citet{2015IBVS.6126....1A} obtained resolved UBVRI photometry of the system in 2015. However, they found that the primary has suffered a $3^\text{m}$ gray extinction, while the secondary got $0.7^\text{m}$ brighter in all bands, compared to the measurements of \citet{2001ApJ...556..265W}. Therefore, since this is a high difference in magnitudes, we omitted including this photometry data in our analysis. 
The variability was also observed in the IR regime: \citet{2015IBVS.6143....1S} noted $\approx 1.5^\text{m}$ variability in the J-band over four years.

RW Aur B had two astrometric epochs that deviate significantly from the hypothetical orbit and the other data points. One epoch is a Keck observation in a narrow-band N-band observation (1999-11-16, IHW18 filter), which may have captured some other feature, such as scattered light from the jets, disk or nebula around RW Aur. The other deviating epoch comes from the survey  of \citet{1993AnA...278..129L}.

\section{Single star sample}

In the following we list the single T Tauri stars that are used in Section~\ref{sec:discussion} to compare the distribution of the 10~$\mu$m IR excess of single stars and stars in multiple systems.

\begin{table*}[ht]
\setlength{\tabcolsep}{3pt}
\caption{Single star sample and the $10~\mu$m IR excesses of those.}
\label{table:singlessedparams}
\centering
\begin{tabular}{llll|lll}
\hline\hline
Star &  SpT & \multicolumn{1}{l}{$T_{eff}$} & $A_V$ & $A_V$& Lum. & \multicolumn{1}{l}{IR excess} \\ 
     &      & \multicolumn{1}{l}{(K)}        & lit. & fit  & ($L_\sun$) & \multicolumn{1}{l}{(mag)}  \\
\hline
      V1304 Tau & M1 &   3630 & -- & $\textcolor{black}{0.8}^{+\textcolor{black}{0.8}}_{-\textcolor{black}{0.4}}$ & 0.25 &  1.9 \\[0.4em]
2MASS J04110570+2216313 & M4 &   3200 & -- & $\textcolor{black}{1.1}^{+\textcolor{black}{0.1}}_{-\textcolor{black}{0.2}}$ & 0.10 &  1.7 \\[0.4em]
         LkCa 1 & M4 &   3200 & 1.0 & $\textcolor{black}{1.0}^{+\textcolor{black}{0.3}}_{-\textcolor{black}{0.5}}$ & 0.33 &  1.7 \\[0.4em]
IRAS 04108+2910 & M3 &   3400 & -- & $\textcolor{black}{0.0}^{+\textcolor{black}{5.7}}_{-\textcolor{black}{0.0}}$ & 0.01 &  9.1 \\[0.4em]
         FM Tau & M0 &   3770 & 0.7 & $\textcolor{black}{1.1}^{+\textcolor{black}{3.3}}_{-\textcolor{black}{1.1}}$ & 0.19 &  4.7 \\[0.4em]
         FN Tau & M5 &   2880 & 1.4 & $\textcolor{black}{1.7}^{+\textcolor{black}{3.0}}_{-\textcolor{black}{1.7}}$ & 0.17 &  6.2 \\[0.4em]
    XEST 20-071 & M3 &   3400 & -- & $\textcolor{black}{4.2}^{+\textcolor{black}{1.8}}_{-\textcolor{black}{3.9}}$ & 0.22 &  2.5 \\[0.4em]
         CY Tau & M2 &   3490 & 1.3 & $\textcolor{black}{1.3}^{+\textcolor{black}{0.7}}_{-\textcolor{black}{0.8}}$ & 0.30 &  4.2 \\[0.4em]
        HBC 372 & K5 &   4140 & 0.0 & $\textcolor{black}{0.0}^{+\textcolor{black}{1.1}}_{-\textcolor{black}{0.0}}$ & 0.12 &  2.0 \\[0.4em]
2MASS J04202606+2804089 & M4 &   3200 & -- & $\textcolor{black}{0.3}^{+\textcolor{black}{0.3}}_{-\textcolor{black}{0.2}}$ & 0.09 &  5.2 \\[0.4em]
         DE Tau & M3 &   3400 & 0.6 & $\textcolor{black}{0.8}^{+\textcolor{black}{1.0}}_{-\textcolor{black}{0.8}}$ & 0.49 &  4.7 \\[0.4em]
         DG Tau & K5 &   4140 & 2.2 & $\textcolor{black}{1.1}^{+\textcolor{black}{3.1}}_{-\textcolor{black}{1.1}}$ & 0.56 &  8.1 \\[0.4em]
        HBC 388 & K1 &   4920 & 0.1 & $\textcolor{black}{0.3}^{+\textcolor{black}{0.9}}_{-\textcolor{black}{0.3}}$ & 1.31 &  1.9 \\[0.4em]
         IQ Tau & M1 &   3630 & 1.3 & $\textcolor{black}{1.2}^{+\textcolor{black}{2.7}}_{-\textcolor{black}{1.2}}$ & 0.36 &  5.0 \\[0.4em]
      V1076 Tau & K7 &   3970 & 0.7 & $\textcolor{black}{1.2}^{+\textcolor{black}{0.7}}_{-\textcolor{black}{0.5}}$ & 0.24 &  1.8 \\[0.4em]
RX J0432.8+1735 & M2 &   3490 & -- & $\textcolor{black}{0.7}^{+\textcolor{black}{1.3}}_{-\textcolor{black}{0.7}}$ & 0.16 &  2.7 \\[0.4em]
       V830 Tau & M0 &   3770 & 0.3 & $\textcolor{black}{0.6}^{+\textcolor{black}{0.7}}_{-\textcolor{black}{0.6}}$ & 0.51 &  1.8 \\[0.4em]
IRAS 04303+2240 & M1 &   3630 & 15.1 & $\textcolor{black}{5.2}^{+\textcolor{black}{0.8}}_{-\textcolor{black}{2.2}}$ & 0.03 &  8.4 \\[0.4em]
2MASS J04334171+1750402 & M4 &   3200 & -- & $\textcolor{black}{1.8}^{+\textcolor{black}{0.5}}_{-\textcolor{black}{0.2}}$ & 0.05 &  3.5 \\[0.4em]
         DM Tau & M2 &   3490 & 0.0 & $\textcolor{black}{0.0}^{+\textcolor{black}{1.6}}_{-\textcolor{black}{0.0}}$ & 0.13 &  6.5 \\[0.4em]
         HO Tau & M1 &   3630 & 1.1 & $\textcolor{black}{0.4}^{+\textcolor{black}{1.5}}_{-\textcolor{black}{0.4}}$ & 0.07 &  5.3 \\[0.4em]
         DN Tau & M1 &   3630 & 0.5 & $\textcolor{black}{0.7}^{+\textcolor{black}{0.8}}_{-\textcolor{black}{0.7}}$ & 0.56 &  4.3 \\[0.4em]
2MASS J04355881+2438404 & M3 &   3400 & -- & $\textcolor{black}{2.1}^{+\textcolor{black}{0.2}}_{-\textcolor{black}{0.3}}$ & 0.10 &  1.7 \\[0.4em]
2MASS J04360131+1726120 & M3 &   3400 & -- & $\textcolor{black}{0.7}^{+\textcolor{black}{3.7}}_{-\textcolor{black}{0.7}}$ & 0.03 &  5.7 \\[0.4em]
        LkCa 15 & K5 &   4140 & 0.6 & $\textcolor{black}{1.3}^{+\textcolor{black}{3.0}}_{-\textcolor{black}{1.1}}$ & 0.51 &  4.2 \\[0.4em]
         GO Tau & K5 &   4140 & 1.2 & $\textcolor{black}{1.7}^{+\textcolor{black}{2.1}}_{-\textcolor{black}{1.7}}$ & 0.07 &  5.4 \\[0.4em]
IRAS 04429+1550 & M3 &   3400 & -- & $\textcolor{black}{1.8}^{+\textcolor{black}{0.3}}_{-\textcolor{black}{1.8}}$ & 0.17 &  5.5 \\[0.4em]
2MASS J04505356+2139233 & M2 &   3490 & -- & $\textcolor{black}{1.6}^{+\textcolor{black}{1.6}}_{-\textcolor{black}{1.6}}$ & 0.17 &  1.7 \\[0.4em]
         GM Aur & K6 &   4020 & 1.4 & $\textcolor{black}{1.2}^{+\textcolor{black}{0.8}}_{-\textcolor{black}{1.0}}$ & 0.55 &  3.9 \\[0.4em]
    XEST 26-071 & M4 &   3200 & -- & $\textcolor{black}{1.7}^{+\textcolor{black}{0.2}}_{-\textcolor{black}{0.6}}$ & 0.07 &  4.3 \\[0.4em]
      V1353 Tau & M1 &   3630 & -- & $\textcolor{black}{1.2}^{+\textcolor{black}{0.5}}_{-\textcolor{black}{0.5}}$ & 0.12 &  1.8 \\[0.4em]
2MASS J04590305+3003004 & M2 &   3490 & -- & $\textcolor{black}{0.6}^{+\textcolor{black}{0.3}}_{-\textcolor{black}{0.2}}$ & 0.11 &  4.4 \\[0.4em]
       V836 Tau & K7 &   3970 & 0.6 & $\textcolor{black}{1.4}^{+\textcolor{black}{0.6}}_{-\textcolor{black}{1.0}}$ & 0.29 &  4.3 \\[0.4em]
2MASS J05044139+2509544 & M4 &   3200 & -- & $\textcolor{black}{1.7}^{+\textcolor{black}{1.1}}_{-\textcolor{black}{1.2}}$ & 0.07 &  4.3 \\[0.4em]
2MASS J05122759+2253492 & M3 &   3400 & -- & $\textcolor{black}{1.2}^{+\textcolor{black}{0.6}}_{-\textcolor{black}{0.7}}$ & 0.20 &  3.0 \\[0.4em]
2MASS J05290743+0150319 & M2 &   3490 & 0.3 & $\textcolor{black}{1.1}^{+\textcolor{black}{0.4}}_{-\textcolor{black}{0.3}}$ & 0.04 &  1.7 \\[0.4em]
2MASS J05374702-0020073 & M2 &   3490 & 0.3 & $\textcolor{black}{1.0}^{+\textcolor{black}{0.9}}_{-\textcolor{black}{1.0}}$ & 0.03 &  4.7 \\[0.4em]
2MASS J05465241+0020016 & K9 &   3880 & 3.3 & $\textcolor{black}{3.2}^{+\textcolor{black}{2.3}}_{-\textcolor{black}{1.0}}$ & 0.02 &  5.7 \\[0.4em]
2MASS J05470397+0011143 & M2 &   3490 & 0.9 & $\textcolor{black}{0.6}^{+\textcolor{black}{0.2}}_{-\textcolor{black}{0.2}}$ & 0.01 &  6.8 \\[0.4em]
[SHB2004] Trumpler 37 24-1736 & M1 &   3630 & 1.0 & $\textcolor{black}{3.0}^{+\textcolor{black}{1.1}}_{-\textcolor{black}{0.4}}$ & 0.01 &  3.4 \\[0.4em]

\hline  
\end{tabular}
\end{table*}

\end{document}